\documentclass{article}
\usepackage{slashed,verbatim}
\usepackage[numbers,sort&compress]{natbib}
\usepackage{amsmath}
 
\renewcommand \thesubsection{\Roman{section}.\Alph{subsection}}

\newcommand{\Rmnum}[1]{\expandafter\@slowromancap\romannumeral #1@}
\makeatother
\usepackage{amssymb}
\usepackage{float}
\usepackage{comment}
\usepackage{appendix}
\usepackage{graphicx}
\usepackage{subcaption}
\usepackage{dcolumn}
\usepackage{bm}
\usepackage[colorlinks=true,linktocpage=true,
linkcolor=blue,citecolor=blue]{hyperref}
\usepackage[a4paper]{geometry}
\newcommand{\nd}{\noindent}
\newcommand{\be}{\begin{eqnarray}}
\newcommand{\ee}{\end{eqnarray}}
\newcommand{\nn}{\nonumber}

\begin{document}

\large
\title{\bf{Seebeck effect in a thermal QCD medium in the presence of strong 
magnetic field}}
\author{Debarshi Dey\footnote{ddey@ph.iitr.ac.in}~~and~~Binoy Krishna
Patra\footnote{binoy@ph.iitr.ac.in}\vspace{0.1in}\\
Department of Physics,\\
Indian Institute of Technology Roorkee, Roorkee 247667, India}
\maketitle
\begin{abstract}
The strongly interacting partonic medium created 
post ultrarelativistic heavy ion collision experiments exhibits 
a significant temperature-gradient between the central and peripheral 
regions of the collisions, which in turn, is capable of inducing an electric 
field in the medium; a phenomenon known as Seebeck effect.
The effect is quantified by the magnitude of the induced electric 
field per unit temperature-gradient - the Seebeck coefficient ($S$).
We study the coefficient, $S$ with the help of the relativistic Boltzmann 
transport equation in relaxation-time approximation, as a function of temperature 
($T$) and chemical potential ($\mu$), wherein we find that with
current quark masses, the magnitude of $S$ for individual quark 
flavours as well as that for the partonic medium decreases with $T$
and increases with $\mu$, with the electric charge of the flavour deciding the sign of $S$. 
The emergence of a strong
magnetic field ($B$) in the non-central collisions at heavy-ion collider 
experiments motivates us to study the effect of $B$ on the Seebeck effect. The
strong $B$ affects $S$ in multifold ways, via : 
a) modification of phase-space due to the dimensional reduction, b) 
dispersion relation in lowest Landau level (occupation probability), and 
c) relaxation-time. We find that a strong $B$ not only decreases the 
magnitudes of $S$'s of individual species, it also flips their signs. This leads to a faster reduction of the magnitude of $S$ of the medium than its counterpart at $B=0$. 
We then explore how the interactions among partons described in 
perturbative thermal QCD in the quasiparticle framework affect 
the Seebeck effect. The interactions indeed affect
the coefficient drastically. For example, even in strong B, there is no
more a flip of the sign of $S$ for individual species and the magnitudes 
of $S$ of individual species as well as that of the medium get enhanced in 
comparison with the current quark mass description at either $B=0$ or $B \neq 0$.
\end{abstract}


\section{Introduction}
The formation of a Quark-Gluon Plasma 
(QGP) under extreme temperatures and/or 
chemical potentials and its subsequent
confinement into interacting hadrons has 
been an intense area of research for more
than three decades. Ultra relativistic Heavy-Ion Collisions (UHRICs) at the 
CERN Super Proton Synchrotron (SPS), Brookhaven National Laboratory 
Relativistic Heavy Ion Collider (RHIC), and Large Hadron Collider
(LHC) accelerators, reach center of mass energies that
are much larger than the critical energy density required
for a transition from hadrons to QGP as
predicted by lattice QCD calculations\cite{Karsh:NuclPhysB605'2001}. Experimental data for hadrons of low to medium 
transverse momenta from the RHIC appeared to quantitatively agree with theoretical results obtained from the macroscopic description of QGP using ideal fluid dynamics, indicating that the 
viscosity of the matter created in the 
early stages post heavy ion collisions is 
small, thus establishing the nature of flow of 
QGP as that of a perfect fluid\cite{KolbeQGP3,Romatschke:PRL99'2007,Schenke:PRC82'2010,PRL106'2011}. Since then,
understanding the
strongly interacting medium formed under
extreme temperatures via studying its
transport coefficients has been a relevant 
and a challenging task. 
In an amazing theoretical discovery, 
Kovtun, Son and Starinets conjectured that
all substances have the value of the ratio 
$\eta/s=1/4\pi$ (in units with $\hbar=k_B=c=1$) 
as the lower limit\cite{Kovtun:PRL95'2005}. 
The smallness of this ratio indeed helped 
to explain the flow data\cite{Heinz:AnnRevNuclPartSci63'2013}. 
There are theoretical 
results that indicate that the ratio of 
bulk viscosity to entropy density  $\zeta/s$ 
may attain a maximum value in the vicinity 
of phase transition, in agreement with lattice 
QCD simulations\cite{Dobado:PRD86'2012,Sasaki:PRD79'2009,Sasaki:NuclPhys62'2010,Karsch:PLB217'2008}. The effect of thermal 
conductivity on the medium has also been 
studied, specifically in relation to the 
determination of the critical point in the 
QCD phase diagram\cite{Kapusta:PRC86'2012}. 
Several methods have been employed in the 
evaluation and study of these transport 
coefficients, which include perturbative QCD, different effective models, etc\cite{Prakash:PR27'1993,Ghosh:PRC90'2014,Kadam:NuclPhysA934'2015}.

Two ultrarelativistic highly charged ions colliding with a 
finite impact parameter can give rise to 
large magnetic fields\cite{Tuchin:AdvHEP'2013}. 
These fields can be as large as $eB\sim10^{-1}
m_{\pi}^2\,(\simeq 10^{17}$ Gauss) for SPS 
energies, $eB\sim m_{\pi}^2$ for RHIC energies 
and $eB\sim 15m_{\pi}^2$ for LHC energies\cite{Sokov:IntJModPhysA24'2009}. The created 
magnetic field was believed to be strong for 
a very short span of time ($\sim$0.2 fm for 
RHIC energies) whereafter it decays very fast\cite{Kharzeev:NuclPhysA803'2008,Asakawa:PRC81'2010}. However, 
it was later pointed out\cite{Tuchin:PRC82'2010,Tuchin:PRC83'2011} 
that owing to a finite electrical conductivity, $\sigma_{el}$ of the plasma,
the magnetic field does not decay very rapidly and hence 
contributes non trivially towards the 
evolution of the medium~\cite{Rath:PRD100'2019}. As such, the effect of magnetic field 
on the transport coefficients also needs to be investigated~\cite{Rath:PRD100'2019}. 

In this work, the thermoelectric behaviour 
of the QGP medium is analyzed via the 
relevant transport coefficient, {\em viz.}, 
the Seebeck coefficient. The first of 
such effects was discovered by T.J. 
Seebeck in 1821 wherein he showed that 
an electromotive force was generated 
on heating the junction between two 
dissimilar metals\cite{Goldsmid121}. 
This phenomenon of conversion of a 
temperature-gradient in a conducting 
medium into an electric current is termed 
as the Seebeck effect and depends on the
bulk properties of the medium involved. 
When a temperature-gradient is established 
in a conducting medium, the more energetic 
 charge carriers diffuse from the region of 
 higher temperature to the region of lower 
 temperature, resulting in the creation of 
 an electric field. The diffusion stops when 
 the created electric field becomes strong 
 enough to impede the further flow of charges. 
 The Seebeck coefficient is defined as the 
 electric field (magnitude) generated in a conducting 
 medium per unit temperature-gradient when 
 the electric current is set to zero\cite{Callen1960,Scheidemantel:PRB68'2003}, 
{\em i.e.} $\vec{E}=S\,\vec{\nabla}T$ ($S$ is the Seebeck coefficient). 
Conventionally, the Seebeck coefficient is 
taken to be positive if the thermoelectric 
current flows from the hotter end to the 
colder end. Thus, the Seebeck coefficient is 
positive for positive charge carriers and 
negative for negative charge carriers, its 
magnitude being very low for metals (only a 
few micro volts per degree Kelvin temperature-gradient) whereas much 
higher for semiconductors (typically a few 
hundred micro volts per degree Kelvin temperature-gradient)\cite
{Ioffe'1957}. It is to be noted that while for 
condensed matter systems, a temperature 
gradient is sufficient to give rise to an 
induced current, this is not the case with a 
system such as the electron-positron plasma 
or the QGP. This is because in a condensed 
matter system, the ions are stationary and 
the majority charge carriers are responsible 
for conduction of electric current. However, 
in a medium consisting of mobile charged 
particles and antiparticles, a temperature 
gradient will cause them to diffuse in the 
same direction, giving rise to equal and 
opposite currents, which cancel. In such 
media, therefore, in addition to a 
temperature-gradient, a finite chemical 
potential is also required for a net induced 
current to exist. Thermoelectric properties 
have been an extensive area of investigation 
in the field of condensed matter physics over 
the past three decades. Some of the notable 
works include the study of the Seebeck effect 
in superconductors\cite{AoArxiv,Matusiak:PRB97'2018,Hooda:EPL121'2018,Choiniere:PRX7'2017,Gaudart:PSS2185'2008}, Seebeck effect in the graphene 
superconductor junction\cite{Wysokinski:JAP113'2013}, 
electric and thermoelectric transport 
properties of correlated quantum dots coupled 
to superconducting electrode\cite{Wojcik:PRB89'2014}, transport 
coefficients of high 
temperature cuprates\cite{Seo:PRB90'2014}, 
thermoelectric properties of a ferromagnet-
superconductor hybrid junction\cite{Dutta:PRB96'2017}, 
Seebeck coefficient in low 
dimensional correlated organic metals\cite{Shahbazi:PRB94'2016}, etc. 

The deconfined  hot
QCD medium created post heavy ion collisions can possess a 
significant temperature-gradient between 
the central and peripheral regions of the 
collisions. Majority of collisions in such 
experiments being non-central, a strong 
magnetic field perpendicular to the reaction 
plane is also expected to be created
and could be sustained by a finite electrical 
conductivity of the medium. Study of thermoelectric properties in the 
context of heavy ion collisions is still 
uncharted territory, apart from a 
single paper\cite{Bhatt:PRD99'2018}, 
wherein the authors calculated the Seebeck 
coefficient of a baryon rich hot hadronic 
gas with zero meson chemical potential, 
modelled by the Hadron Resonance Gas (HRG) 
model at chemical freeze-out\cite{Braun:WS'2004,Andronic:NuclPhysA772'2006} 
with a resonance mass cut-off of 
$2.25$ GeV. However, in the present work, we wish to
do the aforesaid investigation on a color deconfined medium of quarks 
and gluons. In addition, we 
also explore the effects of a strong magnetic field and quasiparticle 
description of the medium constituents, on the Seebeck effect, where the quasiparticle/effective
masses of the partons ({\em mainly} quarks)
are evaluated from perturbative thermal QCD up to one-loop. The noteworthy differences from a hadronic medium are two-fold: 
i) The degrees of freedom are more fundamental, {\em i.e.}
the elementary quarks and gluons instead of mesons and baryons. 
ii) The system is relativistic, {\em i.e.} $m_f\ll T$ ($m_f$ refers to the 
mass of $f^{\mbox{th}}$ flavour).

The paper is organised as follows: In Section II, we discuss 
the thermoelectric effect in a thermal QCD medium in a kinetic
theory approach. In subsection, II. A, we will discuss the Relativistic 
Boltzmann Transport Equation (RBTE) in the relaxation-time approximation,
where we will derive the relaxation-time fora thermal QCD medium with a finite
chemical potential. Then in II. B and II. C, we  quantify the effect 
by the Seebeck coefficients with the current quark masses of the partons
in the absence and presence of a strong magnetic field, respectively.
In Section III, we calculate the same, taking into account
the interactions in the medium by perturbative thermal QCD in
a strong magnetic field, which in turn generates (quasiparticle) 
masses for the partons. In Section IV, the conclusions are drawn and the 
results are summarized.

\section{Seebeck effect in hot partonic medium with current quark masses}
In this section, we construct a general framework for studying the 
thermoelectric effect for a hot partonic medium and then use the
framework to estimate 
the Seebeck coefficient for the individual species as well as 
for the composite medium. Then we study the effects of strong magnetic 
field on the aforesaid study of the thermoelectric effect.
\subsection{Boltzman transport equation and relaxation time}
In this subsection, we will begin with the relativistic Boltzmann transport 
equation (RBTE) to estimate the response of a thermal QCD mediun
to an external electric field. The rate of change of the distribution function 
of the $i^{\mbox{th}}$ flavour 
possessing quark chemical potential $\mu_i$, immersed in a thermal medium of 
quarks and gluons at temperature $T$ is given by 
RBTE
\begin{equation}
p^{\mu}\frac{\partial f_i(x,p)}{\partial 
x^{\mu}}+q_iF^{\rho \sigma}p_{\sigma}\
\frac{\partial f_i(x,p)}{\partial p^{\rho}}
=C[f_i(x,p)],\label{1}
\end{equation}
where $C[f_i(x,p)]$ is the collision term 
and $F^{\rho \sigma}$ is the electromagnetic 
field strength tensor. Since the treatment of 
the collision term using quantum scattering 
theory is a very cumbersome problem, we resort 
to the relaxation-time approximation, which is valid when the deviation 
($\delta f_i$) is much smaller than the original system, {\em i.e.} 
$\delta f_i \ll f_i^{\rm iso}$, where the original distribution $f_i^{iso}$ 
is assumed as the equilibrium (isotropic) distribution function.  In this 
approximation, the collision term is given by
\begin{equation}
C[f_i(x,p)]\simeq -\frac{p_{\nu}u^{\nu}}
{\tau_i}\delta f_i\label{2},
\end{equation}
where $u^{\nu}=(1,0,0,0)$, is the 4-velocity 
of the medium in the local rest frame and $\tau_i$ 
is the relaxation time for quarks in a hot partonic medium.
It is known that for a thermal medium of quarks and antiquarks,
in addition to a temperature gradient, a finite chemical potential is also 
required for a net thermoelectric current to exist. This motivates us 
to evaluate the relaxation time for a thermal QCD medium with a finite 
chemical potential.

Let us first calculate the momentum transport coefficient, {\em namely}
the shear viscosity ($\eta$), which, in turn, gives the relaxation-time. 
This can be understood physically: $\eta$ is a transport 
coefficient which measures how efficiently the momentum could be transported 
across the layers and hence dimensionally its inverse tells how
quick the perturbed system comes back into the (originally) equilibrated
system. We shall start with a pure gluon medium and then 
include quarks in the medium. The gluon-gluon interaction plays the dominant 
role in bringing the perturbed medium back to the original system.
Thus, the relevant scattering process is the gluon-gluon scattering.
In the leading-order ($g^2$) of coupling ($g$), three gluon vertex
gives the three - $s$, $t$ and $u$ channel diagrams
\begin{figure}[H]
\begin{subfigure}{0.32\textwidth}
\includegraphics[width=0.75\textwidth]{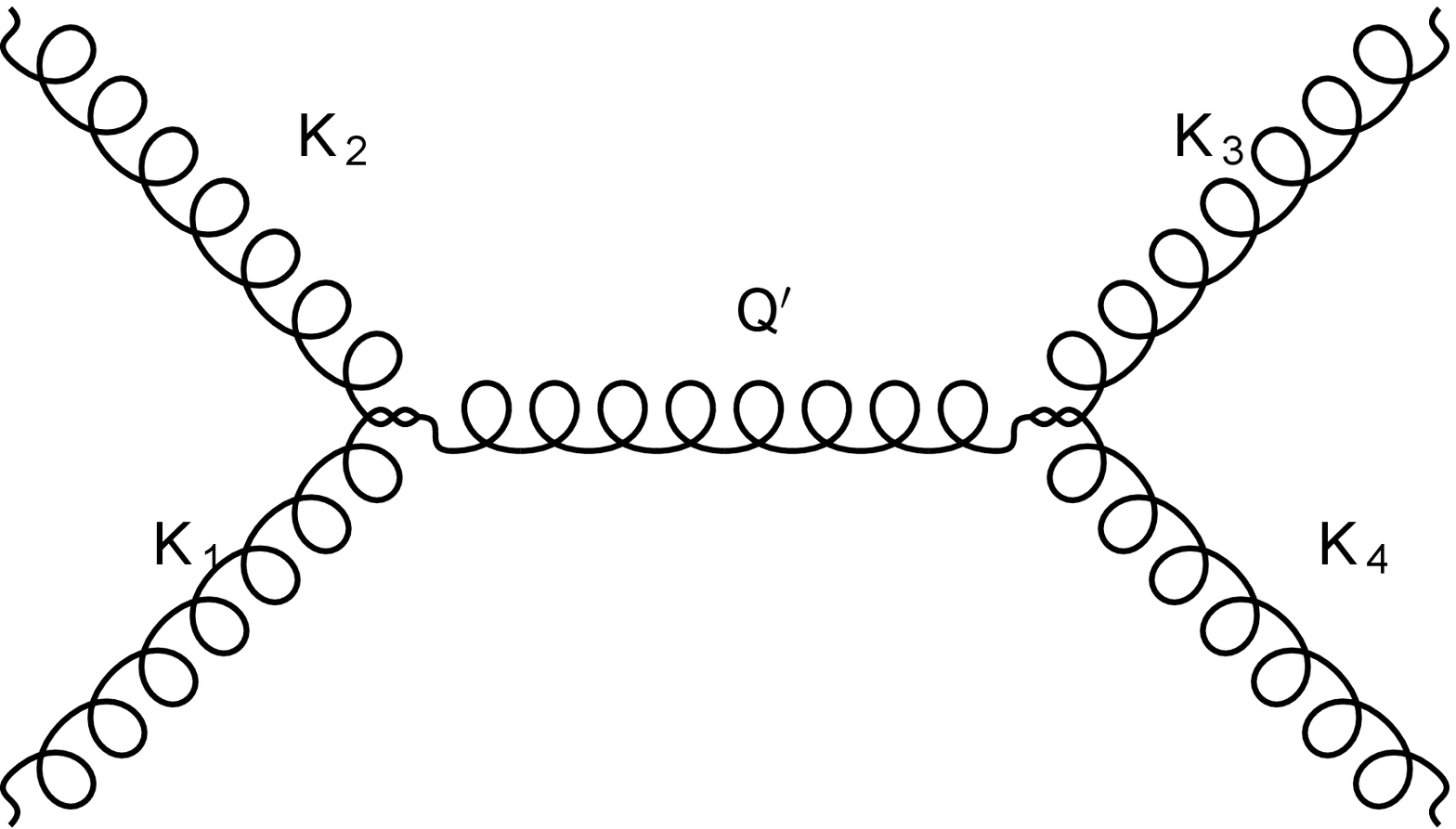}
\caption{}\label{s channel}
\end{subfigure}
\hspace*{\fill}
\begin{subfigure}{0.22\textwidth}
\includegraphics[width=0.65\textwidth]{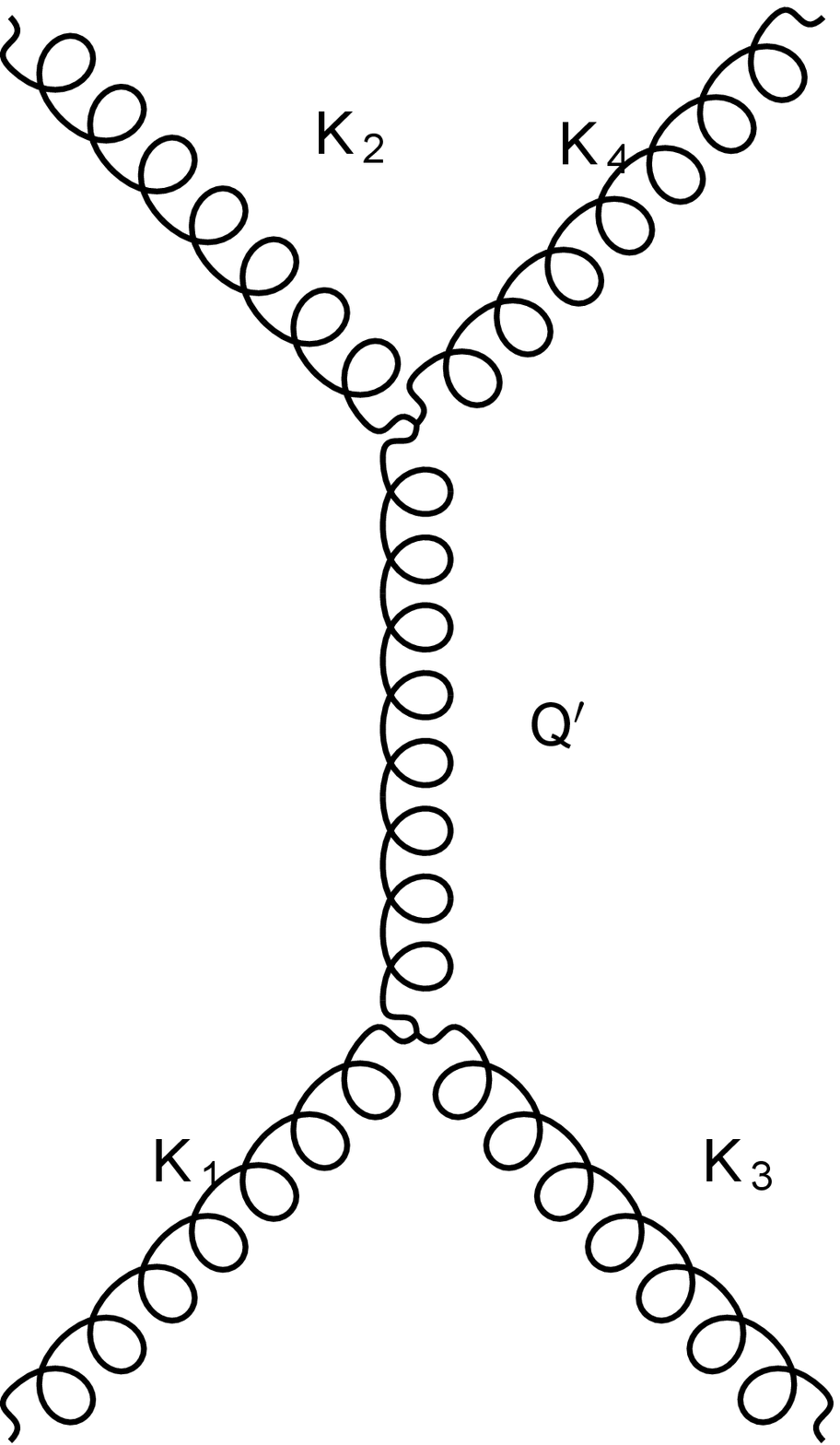}
\caption{}\label{t channel}
\end{subfigure}
\hspace*{\fill}
\begin{subfigure}{0.32\textwidth}
\includegraphics[width=0.75\textwidth]{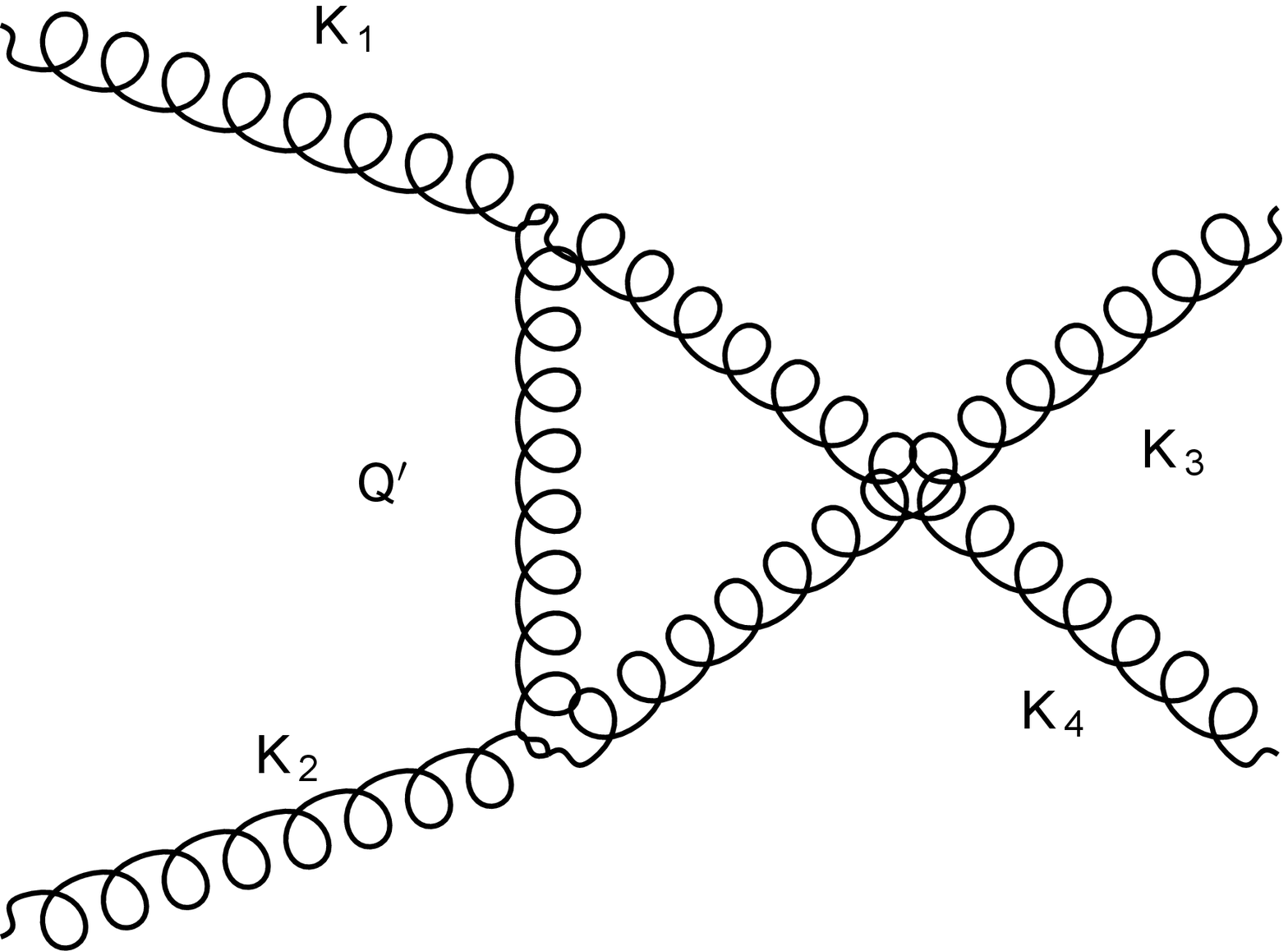}
\caption{}\label{u channel}
\end{subfigure}
\caption{$s$ (left), $t$ (middle) and $u$ (right) channel diagram
}\label{Fig1}
\end{figure}
and the fourth diagram give the four-gluon vertex diagram.
\begin{figure}[H]
        \includegraphics[scale=0.3]{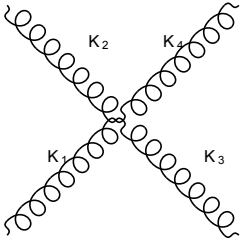}
        \caption{Four-gluon vertex}\label{fgv}
\end{figure}
We have thus calculated the matrix element of all four diagrams 
in terms of Mandelstam variables in Appendix A.
\begin{align}
\overline{|\mathcal{M}|^2}=\frac{9}{2}g^4\left[3-\frac{ut}{s^2}-\frac{us}{t^2}-\frac{st}{u^2}\right].\label{A9}
\end{align}
In the case of near forward scattering, $t$ is much smaller than $s$, so
the above matrix element squared is simplified into the form (Appendix B.1)
\begin{equation}
\overline{|\mathcal{M}|^2}\simeq\frac{9}{2}g^4\frac{s^2}{t^2},\label{D2}
\end{equation}
which, however, results in a singularity in the differential cross-section 
in the forward direction as
\be
\frac{\mbox{d}\sigma}{\mbox{d}\Omega}\sim 
\frac{1}{\sin^4\theta/2},
\ee
where, $\theta$ is the scattering angle in the centre of mass (cm) frame.\\

\nd The singularity appears due to the fact that the gluons are massless.  
The above singularity in the cross-section of ${\rm gg} \rightarrow {\rm gg}$
process in vacuum could be circumvented by the resummed gluon propagator
in thermal medium.\\ 
\nd We consider a frame of reference where the momenta of the incoming gluons are designated as $P(p^0,\vec{p})$ and $P_1(p_1^0,\vec{p_1})$ while that of outgoing (scattered) gluons as $P^{\prime}(p^{\prime\,0},\vec{p^{\prime}})$ and $P_1^{\prime}(p_1^{\prime\,0},\vec{p_1}^{\prime})$, with $p^0\equiv E$, $p_1^0\equiv E_1$ and so on. It will be easier kinematically to study the gluon-gluon scattering
in the forward direction in terms of new four-momenta, $Q (q_0,\vec{q})$,
$K (k_0,\vec{k}) $ and $K^{\prime} (k_0^\prime, \vec{k^\prime})$. These
new momenta are realted to the incmoning and outgoing momenta as
\begin{align}
P&= K+Q/2\,, \qquad \qquad \qquad P_1=K^{\prime}-Q/2 \,\nonumber\\[0.8em]
P^{\prime}&= K-Q/2\,,\qquad \qquad \qquad P_1^{\prime}= K^{\prime}+Q/2.\label{c}
\end{align}
Now,the Mandelstam variables 
can be expressed in terms of the new momenta as
\begin{align}
t&=(P-P^{\prime})^2\nn\\
&=-q^2(1-x^2),\label{F}\\
s&=(P+P_1)^2\nn\\
&\simeq 2kk^{\prime}(1-x^2)(1-\mbox{cos}\,\phi),\label{G}
\end{align}
where, $x=\mbox{cos}\,\alpha =q_0/q$. The angle between 
$\vec{q}$ and $\vec{k}$ is $\alpha$ and the angle between 
the $\vec{p}$-$\vec{p^{\prime}}$ and $\vec{p_1}$-$\vec{p_1^{\prime}}$
planes is $\phi$. 

Thus, substituting $s$ and $t$ in eq.\eqref{D2}, the matrix element squared
can be written as 
\begin{equation}
\overline{|\mathcal{M}|^2}\simeq 18g^4\frac{k^2\,k^{\prime\,2}}{q^4}(1-\mbox{cos}\,\phi)^2.\label{G1}
\end{equation}
To have a consistent perturbative expansion at finite temperature, 
resummation of the gluon propagator is carried out. 
At finite temperature, this propagator gets decomposed into the longitudinal 
($\Delta_L$) and transverse ($\Delta_T$) components with respect to the spatial 
3-momentum ($\vec{q}$) of gluons as: 
\begin{equation}
D_{\mu \nu}(P)= P^T_{\mu \nu}\Delta_T + P^L_{\mu \nu} \frac{Q^2}{q^{2}} 
\Delta_L,\label{PROP} 
\end{equation}
where $\Delta_L$ and $\Delta_T$ are given by: 
\begin{align}
\Delta_L(q_0,q)&=\frac{-1}{q^2+2m_{\rm gT}^2(1-\frac{x}{2}\,\mbox{ln}
	\frac{x+1}{x-1})}\nn\\[0.8em]
\Delta_T(q_0,q)&=\frac{-1}{q_0^2-q^2-m_{\rm gT}^2\left[x^2+
	\frac{x(1-x^2)}{2}\,\mbox{ln}\frac{x+1}{x-1}\right]},\label{I}
\end{align} 
where, $x=\frac{q_0}{q}.$

As a result, the longitudinal component of the propagator in the static limit 
(${\rm q}_0 \rightarrow 0$) manifests the gluons to
acquire an effective mass, $m_{\rm gT}$. 
\begin{equation}
\Delta_L (0, \vec{q})=\frac{-1}{q^{2}+2 m_{{}_{gT}}^2}.
\end{equation}
The effective mass, also known as the thermal mass arises due to
the temperature and screens the infrared singularity. This
is known as Debye screening, which screens the long range
electrostatic fields. On the contrary, the transverse component in the
static limit, at
first sight, does not show the generation of mass, {\em like in the
	longitudinal component}. The singular form of $\Delta_T$ in the static limit
is given by
\begin{eqnarray}
\Delta_T (0, q)=\frac{1}{q^{\,2}}\label{I1},
\end{eqnarray}
implying no screening of the magnetostatic fields. However, if the
leading term in $x$ (=$q_0/|\vec{q}|$) is retained, then the form of $\Delta_T$
manifests a dynamical screening with a cut-off frequency, $\omega_c
={(\pi m_{{}_{gT}}^2 x})^{1/2}$ as
\begin{eqnarray}
\Delta_T (q_0, \vec{q}) \simeq \frac{1}{q^{2} - \frac{i}{2} \pi
	m^2_{gT} \,x}.
\label{deltaT}
\end{eqnarray}
Thus the transverse component of the gluon propagator will now be able to
screen dynamically the infrared singularities to make the cross-sections 
finite, which would otherwise diverge in the bare perturbation theory.

Now we are in a position to calculate the averaged matrix element squared 
for the process $gg \longrightarrow gg$ in near-forward scattering \eqref{G1}
with the resummed gluon propagator \eqref{PROP}. The resummation in the 
matrix element can be effectively introduced by the current-current
interaction mediated by the exchange of a resummed gluon. Thus, the
matrix element is proportional to the current-current correlation  
\begin{eqnarray}
\mathcal{M} &\propto & J^1_\mu D^{\mu\nu}J^2_{\nu} \nonumber\\
& \propto & J^1_L \Delta_L(q_0,q) J^2_L+
\vec{J}^1_T \cdot \vec{J}^2_T \Delta_T(q_0,q).\label{G2}
\end{eqnarray}
The currents, $J^1$, $J^2$ are related to the sources of incoming
and outgoing beam of particles, which, in the limit of 
small momentum transfer ($\vec{q}$), yields into~\cite{Heiselberg:PRD48'1993}:
\begin{equation}
J^1_{\mu}=g\lambda_{\alpha}^1 P_\mu\,,\qquad J^2_{\mu}=g\lambda_{\alpha}^2
P_{1\,\mu},
\end{equation}
where the $\lambda$'s are the Gell-Mann matrices. Thus, the temporal
and spatial components are read off as
\begin{align}
J^1_0&=g\lambda_{\alpha}^1 E \,, \,\quad \vec{ J}^{\,1}_T=g\lambda_{\alpha}^1 E 
\, \vec{v}_{T}\,\nn \\
J^2_0&=g\lambda_{\alpha}^2 E_1,\, \quad \vec{J}^{\,2}_T=g\lambda_{\alpha}^2 
E_1\vec{v_1}_{T}.
\end{align}
The longitudinal components of currents ($J_L$'s) are obtained from $J_0$'s 
by the constraint of the current 
conservation and $|\vec{v}_{T}|=|\vec{v_1}_{T}|=\sqrt{1-x^2}$.

Thus, the matrix element in eq.\eqref{G2} becomes
\begin{equation}
\mathcal{M} = A (k,k^\prime) \left( \Delta_L(q_0,q)+(1-x^2)\mbox{cos}\,\phi 
\Delta_T(q_0,q) \right).\label{G3}
\end{equation}
The function, $A(k,k^\prime)$ is fixed by the fact that in the limit of
vanishing temperature, the matrix element 
squared with the resummed propagator should reduce to its vacuum counterpart. This implies that in the limit of 
vanishing of thermal mass ($m_{\rm gT} \rightarrow 0$), the resummed 
matrix element \eqref{G3} goes to its vacuum result \eqref{G1}. 
Therefore, the matrix element squared \eqref{G3} is thus obtained (Appendix B.2).
\begin{eqnarray}
\overline{|\mathcal{M}|^2}=18g^4k^2k^{\prime\, 2}|\Delta_L(q_0,q)+(1-x^2)
\mbox{cos}\,\phi\, \Delta_T(q_0,q)|^2.\label{H}
\end{eqnarray}

Now we can calculate the collision integral in~\eqref{1} by the 
matrix element for ${\rm gg} \rightarrow {\rm gg}$ scattering, 
\begin{eqnarray}
C[f] &=& \frac{\nu_g}{2 E_{\vec{p}}}\int 
\frac{d^3 p_1}{{(2\pi)}^3 2 E_{\vec{p}_1}}
\frac{d^3 p^\prime}{{(2\pi)}^3 2 E_{\vec{p}^\prime}}
\frac{d^3 p_1^\prime}{{(2\pi)}^3 2 E_{\vec{p}_1^\prime}} 
(2\pi)^4\delta^{(4)}(P+P_1-P^{\prime}-P_1^{\prime}) \nonumber\\
&&\left[f^{\prime}f_1^{\prime}(1+f)(1+f_1)-ff_1(1+f^{\prime})(1+f_1^{\prime})
\right]\overline{|\mathcal{M}|^2},\label{B}
\end{eqnarray}
where, $\nu_g$ is the sum over color and spin degrees of freedom of target gluons.

For a plasma with a local, time independent flow velocity $\vec{u}$, the 
equilibrium distribution function is given by in the Landau frame
\begin{equation}
f^{(0)}=\left(\mbox{exp}\left[\beta(E-\vec{p}.\vec{u})\right]-1\right)^{-1}.\label{J}
\end{equation}
For a non-uniform fluid velocity of the form: $u_x \sim u_x(y)$, 
	the distribution function for the infinitesimally perturbed system 
	becomes
\begin{align}
f&=f^{(0)}+\delta f\nonumber \\
&=f^{(0)}+\frac{\partial f^{(0)}}{\partial E}\,\Gamma(p)
\frac{\partial u_x}{\partial y}.\label{K}
\end{align}
The deviation from equilibrium $\delta f$ is conveniently written by
$\Gamma(p)$, which also satisfies the transport equation. This yields the 
condition for the energy-momentum conservation: 
\begin{equation}
\frac{f^{(0)}}{1+f^{(0)}}\frac{f_1^{(0)}}{1+f_1^{(0)}}=
\frac{f^{\prime\,(0)}}{1+f^{\prime\,(0)}}\frac{f_1^{\prime\,(0)}}{1+
	f_1^{\prime\,(0)}}.\label{L}
\end{equation}
In the above equation, we use the symbols $f$, $f_1$ etc. to denote 
$f\equiv f(\vec{x},\vec{p},\vec{t})$, $f_1\equiv f(\vec{x_1},\vec{p_1},t)$ etc., respectively.

For uniform $\vec{u}$, we note that $f^{(0)}$ satisfies the Boltzmann 
transport equation, 
\begin{equation}
Df^{(0)}=C[f^{(0)}]=0,
\end{equation}
whereas for the non-uniform velocity ($u_x(y)$), the rate of change
of the distribution function in Boltzman transport equation 
in the leading-order of the velocity gradient yields 
\begin{eqnarray}
Df&=&v_y\frac{\partial f^{(0)}}{\partial y}\nonumber \\
&\simeq& \beta p_x v_y n(p) \left(1+n(p)\right)
\frac{\partial u_x}{\partial y},\label{N}
\end{eqnarray}
where, $n\equiv n(p)=(e^{\beta |p|}-1)^{-1}$. 

The prefactor $f^{\prime}f_1^{\prime}(1+f)(1+f_1)-ff_1
(1+f^{\prime})(1+f_1^{\prime})$ of the collision integral \eqref{B}
with the help of Eqs.\eqref{K} and \eqref{L} comes out to be (details are given in 
Appendix B.3).
\begin{eqnarray}
f^{\prime}f_1^{\prime}(1+f)(1+f_1)-ff_1(1+f^{\prime})(1+f_1^{\prime})&=&
\beta (1+f^{\prime\,(0)})(1+f_1^{\prime\,(0)})f^{(0)}f_1^{(0)}\frac{\partial 
	u_x}{\partial y}\left[\Gamma+\Gamma_1-\Gamma^{\prime}-\Gamma_1^{\prime}
\right] \nonumber\\
&\simeq & \beta (1+n^{\prime})(1+n_1^{\prime})\,n\,n_1
\frac{\partial u_x}{\partial y}\left[\Gamma+\Gamma_1-\Gamma^{\prime}-
\Gamma_1^{\prime}\right], 
\label{prefactor}
\end{eqnarray}
where, $\Gamma\equiv\Gamma(p)$, $\Gamma_1\equiv\Gamma(p_1)$ and so on.
Thus the Eqs.\eqref{N} and \eqref{prefactor} have been used to rewrite the
Boltzman transport equation \eqref{1} as
\begin{align}
\beta p_xv_y=&\frac{\nu_g}{2E}\int \frac{\mbox{d}^3p_1}{(2\pi)^3}\, 
\frac{\mbox{d}^3p^{\prime}}{(2\pi)^3}\,\frac{\mbox{d}^3p_1^{\prime}}{(2\pi)^3}(2\pi)^4\delta^{(4)}
(P+P_1-P^{\prime}-P_1^{\prime}) \overline{|\mathcal{M}|^2}\nn\\
&\times \frac{n_1(1+n^{\prime})(1+n_1^{\prime})}{1+n}\left[\Gamma+
\Gamma_1-\Gamma^{\prime}-\Gamma_1^{\prime}\right],\label{P}
\end{align}
By definition, the $x$-$y$ component of the stress tensor to first-order in
velocity gradient is,
\begin{eqnarray}
S_{xy}&=&\nu_g\int\frac{\mbox{d}^3p}{(2\pi)^3}p_{x}v_{y}
\frac{\partial f^{(0)}}{\partial E}\Gamma\frac{\partial u_x}{\partial y}
\nonumber\\
&=& -\eta\frac{\partial u_x}{\partial y},\label{Q}
\end{eqnarray}
This, in turn, expresses the viscosity as
\begin{equation}
\eta \simeq - \nu_g\int \frac{\mbox{d}^3p}{(2\pi)^3}  
p_{x}v_{y}\frac{\partial n}{\partial E}\,\Gamma.
\label{R}
\end{equation}
Thus, the usage of Boltzaman transport equation (Eq.\eqref{P}) and
the symmetry properties of the collision term 
facilitate to write down the viscosity in terms
of the matrix element (details are given in Appendix B.4)
\begin{align}
\frac{1}{\eta}=&
\Bigg(\frac{\beta}{4}\int \frac{\mbox{d}^3p}{(2\pi)^3}\,\frac{\mbox{d}^3p_1}{(2\pi)^3}\,\frac{\mbox{d}^3p^{\prime}}{(2\pi)^3}\,\frac{\mbox{d}^3p_1^{\prime}}{(2\pi)^3}\, n\,n_1(1+n^{\prime})(1+n_1^{\prime})\,\overline{|\mathcal{M}|^2}\times (2\pi)^4\\
&\times\left. \delta^{(4)}(P+P_1-P^{\prime}-P_1^{\prime})[\Gamma+\Gamma_1-\Gamma^{\prime}-
\Gamma_1^{\prime}]^2\Bigg) \middle/  \left(\int \frac{d^3p}{(2\pi)^3}p_{x}v_{y}
\frac{\partial n}{\partial E}\Gamma\right)^{2}\right..
\label{Y}
\end{align}
The integral in the denominator in $\eta$ becomes a standard 
integral and has been evaluated in Appendix B.5.
\begin{equation}
\int \frac{d^3p}{(2\pi)^3}p_{x}v_{y}
\frac{\partial n}{\partial E}\Gamma
=\frac{4T^5\zeta(5)}{\pi^2}.
\label{Z}
\end{equation}
The numerator in $\eta$ is given by the integral (Appendix B.6):
\begin{align}
N\equiv& \frac{\beta}{4} (2\pi)^4 \int \frac{\mbox{d}^3p}{(2\pi)^3}\,\frac{\mbox{d}^3p_1}{(2\pi)^3}\,\frac{\mbox{d}^3p^{\prime}}{(2\pi)^3}\,\frac{\mbox{d}^3p_1^{\prime}}{(2\pi)^3}\, n\,n_1 
(1+n^{\prime}) (1+n_1^{\prime})~\overline{|\mathcal{M}|^2}\times\nn\\
&\delta^{(4)}(P+P_1-P^{\prime}-P_1^{\prime})[\Gamma+\Gamma_1-
\Gamma^{\prime}-\Gamma_1^{\prime}]^2 \nonumber\\[0.6em]
=&\frac{18 \beta g^4}{32(2\pi)^5}\int k^2\frac{\mbox{d}n}{\mbox{d}k}
\mbox{d}k\int k^{\prime\,2}\frac{\mbox{d}n}{\mbox{d}k^{\prime}}
\mbox{d}k^{\prime}\int \mbox{d}x\int \frac{\mbox{d}\phi}{2\pi}
\int \mbox{d}q\, q^3(qx)^2f(qx)[1+f(qx)]\nonumber\\
&\times W(k,k^{\prime};x,\phi)|\Delta_L(q_0,q)+(1-x^2)\mbox{cos}\,\phi\, 
\Delta_T(q_0,q)|^2~, 
\label{Z4}
\end{align}
which diverges in the static limit ($x=0$) of the transverse component of
the resumed propagator, $\Delta_T$ \eqref{I1}. However, as 
mentioned earlier, the divergence could be circumvented 
by retaining the leading term in $x$ in \eqref{deltaT}. Thus, the numerator
comes out to be finite (shown in Appendix B.6)
\begin{equation}
N=\frac{\pi^3T^7\alpha_s^2}{30}.\label{Z8}
\end{equation}

Therefore, the inverse of the shear viscosity in Eq.\eqref{Y} yields into 
the form
\begin{equation}
\frac{1}{\eta}=\frac{\pi^7\alpha_s^2}{T^3\,480\,(\zeta(5))^2}\,
\mbox{ln}\left(\frac{T}{m_{gT}}\right).\label{Z9}
\end{equation}
where the gluon thermal mass for the pure gluonic medium is 
given by~\cite{Vija:PLB342'1995}
\begin{equation}
{m_{gT}}^2(T)=\frac{g^2T^2}{2}.
\label{Z10}
\end{equation}
Thus, the shear viscosity for a pure gluonic medium becomes 
\begin{equation}
\frac{1}{\eta}=\frac{\pi^7\alpha_s^2}{T^3\,960\,(\zeta(5))^2}\,
\mbox{ln}\left(\frac{1}{2\pi\alpha_s}\right).\label{Z11}
\end{equation}

Let us now calculate the relaxation-time from the Boltzmann transport 
equation, which, in relaxation-time approximation, is given by
(Appendix B.7)
\begin{equation}
p_xv_y=\frac{\Gamma}{\tau}.\label{z11}
\end{equation}
Therefore, in conjunction with the definition of stress tensor, 
we obtain the relaxation-time from Eq.\eqref{R} (the relevant 
integral is given in Appendix B.8)
\begin{eqnarray}
\frac{1}{\tau} \simeq \frac{1.404\,T^4}{\eta} \label{taugeneral}
\end{eqnarray}
\begin{equation}
i.e.,\quad \frac{1}{\tau} \approx4.11\, T \alpha_s^2\,\mbox{ln}\left(\frac{1}{2\pi\alpha_s}\right). 
\label{Z14}
\end{equation}

The preceding discussion can now be easily generalized to include the quarks by
including the relevant processes: ${\rm qg} \rightarrow 
{\rm qg}$, ${\rm qq} \rightarrow {\rm qq}$, in addition to the
aforesaid ${\rm gg} \rightarrow {\rm gg}$ process.
Like earlier, the infra-red singularities are again tamed by the masses 
generated in thermal medium, which will now acquire additional
flavour factor as
\begin{equation}
{m_{gT}}^2(T)=\frac{g^2T^2}{2} \left( 1 +\frac{N_f}{6}\right).
\label{mgt_fl}
\end{equation}

One then finds that the gluon ($g$) and quark ($q$) contributions are simply 
added to yield the final form of the relaxation time. The quark ($\eta_q$) 
and gluon ($\eta_g$) contributions to the viscosity of the medium 
are given by \cite{Bellac:BOOK'1996}
\begin{eqnarray}
\eta_g &=& \frac{\eta}{1+N_f/6}, \\
\eta_q &\simeq& 1.70N_f\,\eta_g,
\end{eqnarray} 
respectively. Thus the total viscosity becomes
\begin{eqnarray}
\eta_{tot} &=& \eta_g+\eta_q \nonumber\\
&=&\frac{\eta}{1+N_f/6}\left(1+1.70\,N_f\right),
\end{eqnarray}
where, $\eta$ is given by Eq.\eqref{Z11}. Hence, the relaxation-time for 
a quark-gluon thermal medium is obtained from Eq.\eqref{taugeneral} as   
\begin{equation}
\frac{1}{\tau (T)} \approx \frac{0.4(N_f+6)}{(N_f+0.6)}~T\alpha_s^2\,
\mbox{ln}\left(\frac{1}{2\pi\alpha_s}\right).
\end{equation}

The case discussed above was limited to the baryonless medium
($\mu_q=0$, $\mu_q$ is the quark chemical potential). Now we wish
to evaluate the above relaxation time for a thermal QCD medium
with a finite chemical potential. Then, in addition to the
temperature,  the dependence of the chemical potential
will also enter into the gluon self-energy, which, in turn, introduces
$T$ and $\mu_q$ dependence into the
above-mentioned longitudinal and transverse components of the propagator.
As a consequence, the gluon thermal mass
will now depend on both $T$ and $\mu_q$, which will ultimately
screen the infrared singularities, resulting the cross sections of the
above processes finite. Thus the relaxation-time will ultimately
be modified by the $m_{\rm gT}$, which, in turn, makes the relaxation-time
($\tau$) temperature and chemical potential dependent.\\

Thus we start with a thermal QCD medium with finite chemical potential and
the thermal mass for gluons for such an medium comes out to be~\cite{Vija:PLB342'1995}
\begin{equation}
{m_{gT}}^2(T,\mu)=\frac{g^2T^2}{2}\left[1+\frac{N_f}{6}\left(1+
\frac{1}{\pi^2} \frac{\sum_i \mu_i^2}{T^2}\right)\right],
\label{mgtchem}
\end{equation}
where $\mu_i$ is the chemical potential for ${\rm i}^{\mbox{th}}$ flavour.
Therefore, proceeding as before, we finally arrive at the expression of 
relaxation-time 
\begin{equation}
\frac{1}{\tau (T,\mu)} = \frac{0.4.(N_f+6)}{(N_f+0.6)}~T\alpha_s^2\,
\mbox{ln}\left[\frac{1}{2\pi\alpha_s}\,\frac{1}{\left( (N_f+1)+
	\frac{3}{\pi^2} \sum_i\frac{\mu_i^2}{T^2}\right)}\right]\label{rt}.
\end{equation}
Here, the strong coupling runs with the temperature and chemical potential 
as~\cite{Haque:PRD87'2013}
\begin{align}
\alpha_{s}(T,\mu_i) &=\frac{g^2(T,\mu_i)}
{4\pi}\nonumber \\
&=\frac{6\pi}{(33-2N_f)\,\mbox{ln}\left
	(\frac{2\pi \sqrt{T^2+\mu_i^2/\pi^2}}{
		\Lambda_{QCD}}\right)}.\label{100}
\end{align}

\subsection{Seebeck coefficient in the absence of magnetic filed}
Thus, once the collision integral (eq.\ref{2}) is known, one can 
in principle obtain the infinitesimal disturbance, $\delta f_i$ from RBTE
(eq.\ref{1}). For the thermoelectric effect, we are interested only in 
the quark distribution for which the equilibrium  
distribution for $i^{\mbox{th}}$ flavour, $f_i$,
is given by
\begin{equation}
f_i=\frac{1}{\exp\left(\frac{w_i-\mu_i}
{T(\vec{r})}\right)+1}.\label{4}
\end{equation}
Suppose an infinitesimal electric field, $\vec{E}$ (= $F^{i0}$ or - $F^{0i}$)
disturbs the above equilibrium distribution in
phase-space infinitesimally and the infinitesimal
disturbance from equilibrium ($\delta f_i$) also satisfies the
RBTE. Thus, the disturbance is obtained by considering 
the $\rho=i$ and $\sigma=0$ components in RBTE (eq.\ref{1}). 
\begin{equation}
\delta f_i=\frac{\vec{p_i}\,\tau_i}
{\omega_i}f_i(1-f_i)\left(-\frac{1}{T^2}
\right)\nabla_{\vec{r}}\,T(\vec{r})+2q_
if_i(1-f_i)\left(\vec{E_i}.\vec{p_i}\right)
\frac{\tau_i}{\omega_i\,T}, \label{8}
\end{equation}
\vspace{1.5mm}
which, in turn, produces the induced four-current 
through the relation
\begin{equation}
j_{\mu}= \sum_{i}q_ig_i\int\frac{d^3\,\mbox{p}}{(2\pi)^3\,\omega_i}
p_{\mu}\left[\delta f_i^q(x,p)-\delta f_i^{\bar{q}}
(x,p)\right], \label{9}
\end{equation}
where $q_i$ and $g_i$ are the charge and degeneracy 
factors of the $i^{ \mbox{th}}$ quark flavour, respectively. 

We thus obtain the spatial-part of the induced four-current, 
{\em i.e.} the induced current density
\begin{multline}
\vec{j_i}= \frac{g_iq_i\tau_i}{2\pi^2}\Bigg[\int_{0}
^{\infty}\mbox{dp}\,\frac{p_i^4}{\omega_i^2(p)}
\left\{f_i (1-f_i)(\omega_i (p)-\mu_i)+\bar{f_i}(1-
\bar{f}_i)(\omega_i (p)+\mu_i)\right\}\left(-
\frac{1}{T^2}\right)\nabla_{\vec{r}}T(\vec{r})\\
+\,2q_i \int_{0}^{\infty}\mbox{dp}\,\frac{p_i^4}{\omega_i^2(p)}\left
\{f_i(1-f_i)+\bar{f}_i (1-\bar{f}_i)\right\}\frac{\vec{E_i}}
{T(\vec{r})}\Bigg].\label{10}
\end{multline} 
The above current density is set equal to zero to yield 
the electric field due to the temperature-gradient
\cite{Nolas:Springer45'2001}. Thus, the relation between the 
temperature-gradient in the coordinate space and the induced
electric field is obtained:
\begin{equation}
	\vec{E_i}=\frac{1}{2Tq_i}\frac
	{\int_{0}^{\infty}\mbox{dp}\,\frac{p_i^4}
	{\omega_i^2(p)}\left\{f_i (1-f_i)(\omega_i (p)-\mu_i)
	-\bar{f_i}(1-\bar{f_i})(\omega_i (p)+\mu_i)
	\right\}}{\int_{0}^{\infty}\mbox{dp}\,
	\frac{p_i^4}{\omega_I^2(p)}\left\{f_i(1-f_i)+
	\bar{f}_i (1-\bar{f}_i)\right\}}\,\nabla_{\vec{r}}\,T(\vec{r}),\label{11}
\end{equation}
For a single species, the degeneracy factor and the relaxation time 
cancel out from the numerator and denominator. Thus, the induced electric 
field can be recast in the form
\begin{equation}
\vec{E}=\frac{1}{2Tq}\frac{I_2}{I_1}\,\nabla_{\vec{r}}T(\vec{r}),\label{12}
\end{equation}
where the integrals $I_1$ and $I_2$ are defined by
\begin{align}
I_1& \equiv \int_{0}^{\infty}\mbox{dp}\,\frac{p^4}{\omega^2(p)}\left\{f(1-f)+\bar{f}(1-\bar{f})\right\}\\
I_2& \equiv \int_{0}^{\infty}\mbox{dp}\,\frac{p^4}{\omega^2(p)}\left\{f(1-f)
(\omega(p)-\mu)-\bar{f}(1-\bar{f})(\omega(p)+\mu)\right\},\label{two}
\end{align}
where the chemical potential ($\mu_i$) for all flavours are taken the
same, {\em i.e.} $\mu_i=\mu.$

Therefore, the coefficient of the temperature-gradient in the above 
relation (eq.\ref{12}) gives the Seebeck coefficient ($S$) for a single species 
\begin{equation}
S=\frac{1}{2Tq}\frac{I_2}{I_1}.
\label{13}
\end{equation}
We compute the coefficient, $S$ as a function of temperature 
at fixed chemical potentials, $\mu_q$ =30 MeV, 40 MeV and 50 
MeV to observe the Seebeck effect in a hot and dense medium. 
As per recent studies, the transition temperature for the transition from 
hadron phase to QGP phase for $2+1$ flavours is $154\pm 9$ 
MeV~\cite{Ding:IJMPE24'2015}. The temperature range considered here, 
is $T=$ 165 MeV- 450 MeV.
  
\vspace{6mm}
\begin{figure}[H]
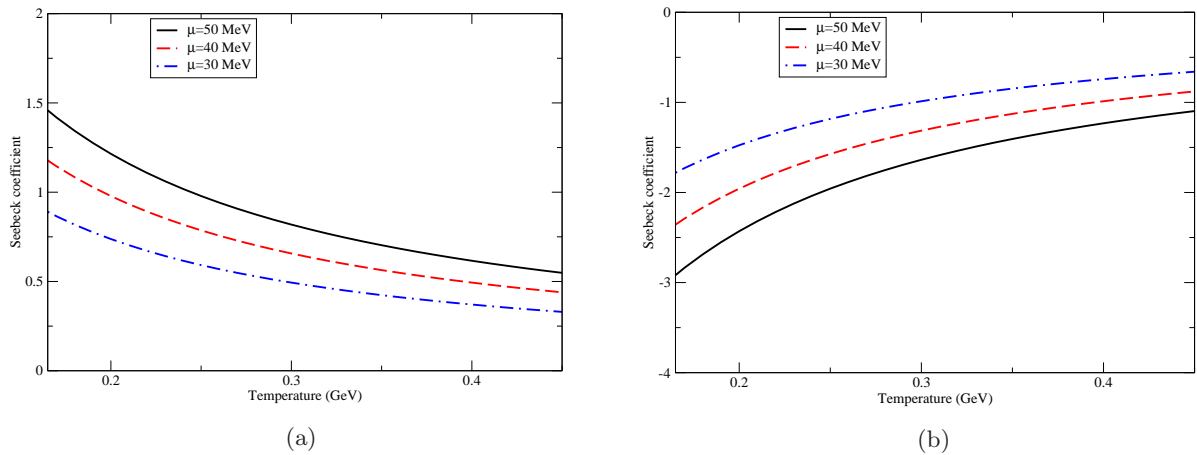

\begin{subfigure}{0.48\textwidth}
\includegraphics[width=0.95\textwidth]{uT.eps}
\caption{}\label{fig1a}
\end{subfigure}
\hspace*{\fill}
\begin{subfigure}{0.48\textwidth}
\includegraphics[width=0.95\textwidth]{dT.eps}
\caption{}\label{fig1b}
\end{subfigure}
\caption{Variation of Seebeck coefficient of 
$u$ (left) and $d$ (right) quarks with 
temperature for different fixed values of 
chemical potentials.}\label{fig1}
\end{figure}
We observe that the Seebeck coefficients (magnitudes) for $u$ and $d$ quark  (Figs. \ref{fig1a} and \ref{fig1b})
decrease with the temperature for a fixed chemical 
potential, which is due to the fact that the net number density, ($n- 
\bar{n}$) (which is proportional to the net charge) decreases with the
temperature for a fixed $\mu$. However, the coefficient is found to 
increase with chemical potential at a given temperature. This is because a larger $\mu$ is indicative of a larger surplus of  particles 
over anti-particles, which, in the case of $u$ quark 
implies a larger abundance of positive over negative charges, 
leading to a larger thermoelectric current and hence a larger $S$. 
However, for the $d$ quark, a larger $\mu_q$ would mean a larger abundance 
of negative charges (particles) over positive charges (anti-particles), 
leading to a more negative value of $S$. Here, the sign of the Seebeck 
coefficient is solely determined by the sign of the 
electric charge the particle carries, because the other
factors in the coefficient- the integrals $I_1$ 
and $I_2$ (eq.\eqref{13}), for both the quarks, are positive 
as can be seen from Fig(\ref{Fig3}) and Fig(\ref{Fig4}). The current quark 
masses of $u$ and $d$ quarks being very close to 
each other leads to almost identical values of the 
$I_1$ and $I_2$ integrals for both the quarks. As 
such, the magnitude of the electric charge of $u$ 
quark being twice that of $d$ quark is directly reflected 
in the magnitude of Seebeck coefficient of $d$ quark 
being almost twice that of the $u$ quark.
\vspace{8mm}
 \begin{figure}[H]
	\centering
	\includegraphics[width=9.5cm,height=9.2cm,keepaspectratio]{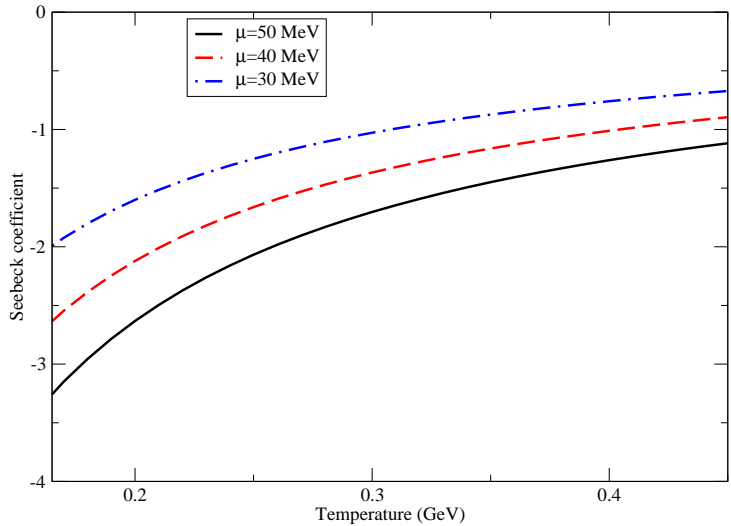}
	\caption{Variation of $s$ quark Seebeck 
coefficient with temperature for different 
fixed values of chemical potentials.}
\label{Fig2}
\end{figure}
The Seebeck coefficient for $S$ quark shows the same characteristics as 
that of $u$ and $d$ quarks. The $I_1$ and $I_2$ integrals are positive for the 
$s$ quark as well. However, its mass 
is almost $100$ times more than that of $u$ or $d$. 
This leads to a larger range of $I_1$ and $I_2$ values 
for the $s$ quark as compared to the $d$ quark. 
However, the ratio $I_1/I_2$ yields values that are 
not too different from that of $d$ quark (Fig (\ref{Fig3}) \& Fig (\ref{Fig4})). 
The electric charge for the $s$ quark is the same as $d$ quark, so they exhibit 
close agreement in the values of respective individual Seebeck coefficients.
\vspace{3mm}
\begin{figure}[H]
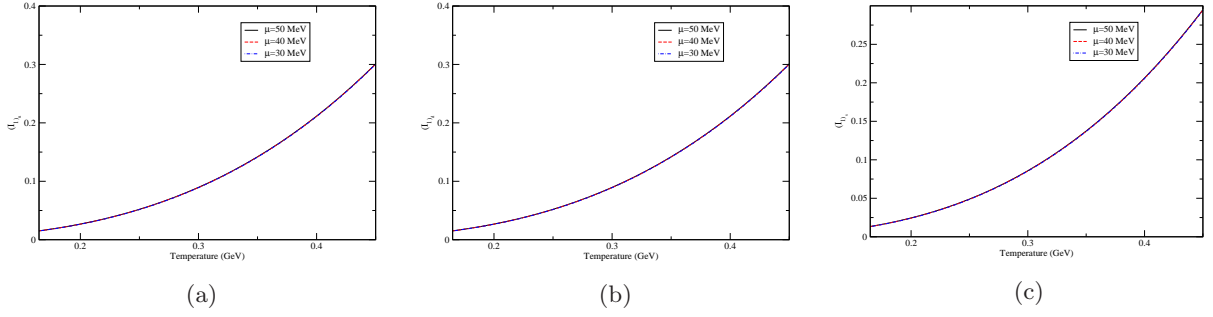

\begin{subfigure}{0.32\textwidth}
\includegraphics[width=0.95\textwidth]{I1unqpm.eps}
\caption{}\label{fig3a}
\end{subfigure}
\hspace*{\fill}
\begin{subfigure}{0.32\textwidth}
\includegraphics[width=0.95\textwidth]{I1dnqpm.eps}
\caption{}\label{fig3b}
\end{subfigure}
\hspace*{\fill}
\begin{subfigure}{0.32\textwidth}
\includegraphics[width=0.95\textwidth]{I1snqpm.eps}
\caption{}\label{fig3c}
\end{subfigure}
\caption{Variation of $I_1$ integrals for $u$ (left), $d$ (middle) and $s$ (right) quarks with temperature for different 
fixed values of chemical potential.}\label{Fig3}
\end{figure}

\begin{figure}[H]
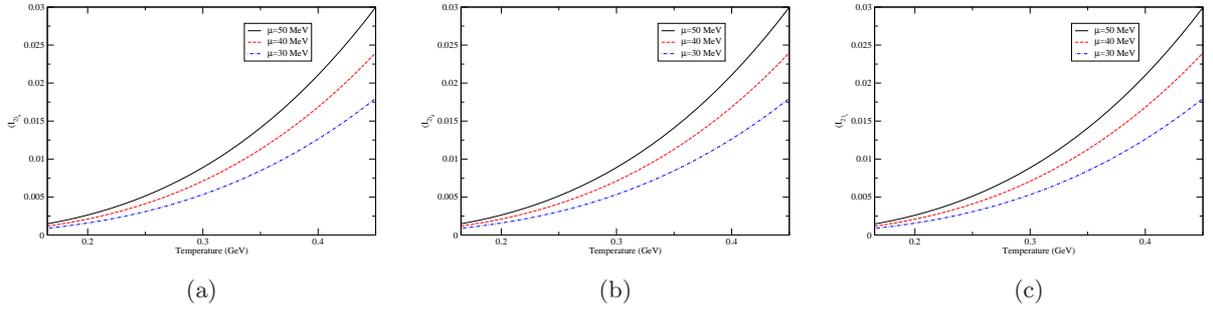

\begin{subfigure}{0.32\textwidth}
\includegraphics[width=0.95\textwidth]{I2unqpm.eps}
\caption{}\label{fig4a}
\end{subfigure}
\hspace*{\fill}
\begin{subfigure}{0.32\textwidth}
\includegraphics[width=0.95\textwidth]{I2dnqpm.eps}
\caption{}\label{fig4b}
\end{subfigure}
\hspace*{\fill}
\begin{subfigure}{0.32\textwidth}
\includegraphics[width=0.95\textwidth]{I2snqpm.eps}
\caption{}\label{fig4c}
\end{subfigure}
\caption{Variation of $I_2$ integrals for $u$ (left), $d$ (middle) and $s$ (right) quarks with temperature for different fixed values of chemical potential.}\label{Fig4}
\end{figure}
	
After having calculated the Seebeck coefficient 
for a thermal medium consisting of a single 
species, we move on to the more realistic 
case of a multi-component system, which in 
our case corresponds to multiple flavours of 
quarks in the QGP. However, gluons being electrically 
neutral, do not contribute to the thermoelectric 
current, therefore, the total electric current in the medium is 
the vector sum of currents due to individual species:
\begin{align}
\vec{J}&=\vec{J_{(1)}}+\vec{J_{(2)}}+\vec{J_
{(3)}}+ \cdots \nonumber\\
&=\left(\frac{q_1^2g_1\tau_1}{T\pi^2}(I_1)_1+\frac{q_2^2g_2\tau_1}{T\pi^2}(I_1)_2+...
\right)\vec{E}-\bigg(\frac{q_1g_1\tau_1}
{2T^2\pi^2}(I_2)_1+\frac{q_2g_2\tau_2}{2T^2
\pi^2}(I_2)_2+...\bigg)\nabla_{\vec{r}}T
(\vec{r}).\label{14}
\end{align}
Setting the total current, $\vec{J}=0$ as earlier, we get the 
induced electric field,
\begin{equation}
\vec{E}=\frac{\sum_{i}\frac{q_ig_i
\tau_i(I_2)_i}{2T}}{\sum_{i}q_i^2g_i\tau_i
(I_1)_i}\nabla_{\vec{r}}T(\vec{r}),
\label{15}
\end{equation}
which yields the Seebeck coefficient for the multi-component medium:
\begin{equation}
S= \frac{1}{2T}\frac{\sum_{i}q_
ig_i\tau_i(I_2)_i}{\sum_{i}q_i^2g_i\tau_
i(I_1)_i}.\label{16}
\end{equation}
All quarks have the same degeneracy factor and their relaxation times
(seen from eq.\eqref{rt}) are also identical for each flavour. 
Hence, the total Seebeck coefficient for the multi-component 
systems can be rewritten as
\begin{equation}
S=\frac{\sum_{i} S_i\,q_i^2(I_1)_i}{\sum_
{i}q_i^2(I_1)_i},\label{17}
\end{equation}
which could be viewed as a weighted
average of the Seebeck coefficients of 
individual species ($S_i$) present in the medium.
In our calculation, we have considered only three 
flavours of quarks, viz: $u$, $d$ and $s$, thus, the explicit expression 
comes out to be:
\begin{equation}
S=\frac{4S_u(I_1)_u+S_d(I_1)_d+S_s(I_1)_s}
{4(I_1)_u+(I_1)_d+(I_1)_s},\label{18}
\end{equation}
where $S_u$, $S_d$, $S_s$ denote the individual 
Seebeck coefficients for the $u$, $d$ and $s$ 
quarks respectively. Likewise, the $I_1$ 
integrals for different flavours are denoted 
by the respective flavour indices.
\vspace{4mm}
 \begin{figure}[H]
	\centering
	\includegraphics[width=9.5cm,height=9.2cm,keepaspectratio]{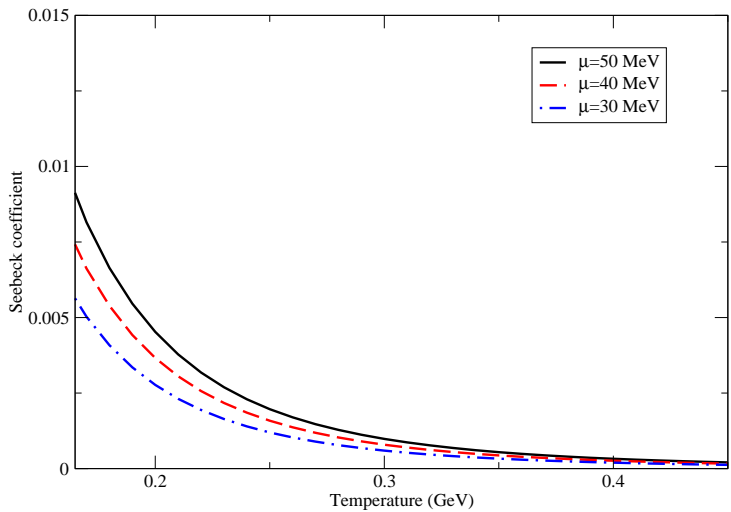}
	\caption{Variation of total Seebeck 
coefficient with temperature for different 
fixed values of chemical potentials.}
\label{fig5}
\end{figure}
As can be seen from Fig(\ref{fig5}), the Seebeck coefficient of the medium is positive and decreases with the temperature. Like earlier for 
single species, it increases with the chemical potential. Although $S_d$ and $S_s$ 
are both negative, the relative magnitudes of $S_u$, $S_d$ and $S_s$ are 
such that  eq.\eqref{18} renders the Seebeck 
coefficient of the medium positive with a small magnitude.

The magnitude of the Seebeck coefficient 
is the magnitude of electric field produced 
in the medium for a unit temperature-gradient. 
Qualitatively, it is a measure of how 
efficiently a medium can convert a 
temperature-gradient into electricity. 
The sign of the Seebeck coefficient expresses 
the direction of the induced field with 
respect to the direction of temperature 
gradient, which is conventionally taken 
to point towards the direction of increasing 
temperature. A positive value of the Seebeck 
coefficient means that the induced field is 
in the direction of the temperature-gradient. 
In the convention mentioned above, this will 
happen when the majority charge carriers are
positively charged.  As expected, individual 
Seebeck coefficient is positive for a 
positively charged species ($u$ quark) and 
negative for a negatively charged species 
($d$ and $s$ quarks). 

\subsection{Seebeck coefficient in the presence of a strong magnetic field}
In the presence of magnetic field, we decompose the quark momentum 
into components longitudinal ($p_L$) and transverse ($p_T$) to the direction
of the magnetic field. Quantum mechanically the energy levels 
of the $i^{\mbox{th}}$ quark flavour get discretized
into Landau levels, so the dispersion relation becomes 
\begin{equation}
\omega_{(i,n)}(p_L)=\sqrt{p_{L}^2+m_
i^2+2n|q_iB|},\label{19} 
\end{equation}
where $n=0,1,2, \cdots $ are quantum numbers 
specifying the Landau levels. It is well 
known that in the strong magnetic field (SMF) 
limit (characterised by $|q_fB|\gg T^2$, where $B$ 
is the magnetic field and $q_f$ is the electric charge 
of the $f^{\mbox{th}}$ flavour), quarks are rarely excited thermally to 
higher Landau levels owing to the large energy 
gap between the levels, which is of the order 
of $\sqrt{|eB|}$\cite{Landau:QM3}. Therefore, they are constrained
to be populated exclusively in the lowest Landau level (n=0), implying that 
the quark 
momentum in the presence of a strong magnetic field is 
purely longitudinal\cite{Gusynin:PLB450'1999,Rath:JHEP12'2017,
Rath:Arxiv}. Taking the magnetic field to be 
in the 3-direction, we identify $p_L$ with 
$p_3$, so the above dispersion relation is simplified
into a relation for a one-dimensional free particle :
\begin{equation}
\omega_{i}(p_3)=\sqrt{p_{3}^2+m_i^2}.\label{20}
\end{equation}
Thus, the equilibrium quark distribution function  in SMF limit
becomes: 
\begin{equation}
f_{i,B}=\frac{1}{e^{\beta (\omega_i-\mu_i)}+1}.\label{23}
\end{equation}
Owing to the quark momentum being purely longitudinal in the presence of a strong magnetic field, the  electromagnetic 
current generated in response to the electric field ($J_3$) is also purely longitudinal.
\begin{equation}
J_{3}= \sum_{i}q_ig_i\int\frac{d^3\,\mbox{p}}
{(2\pi)^3\,\omega_i}p_{3}\left[\delta f_i^q
(\tilde{x},\tilde{p})-\delta f_i^{\bar{q}}
(\tilde{x},\tilde{p})\right],\label{24}
\end{equation}
where, $\tilde{x}=(x_0,0,0,x_3)$ and $\tilde{p}=(p_0,0,0,p_3)$.
In addition, as an artifact of strong magnetic field, 
the density of states in two spatial directions perpendicular to the 
direction of magnetic field becomes $|q_iB|$~\cite{Gusynin:NuclPhysB462'1996,
Bruckmann:PRD96'2017}, {\em i.e.}
\begin{equation}
\int \frac{d^3p}{(2\pi)^3}\rightarrow\frac
{q_iB}{2\pi}\int \frac{dp_3}{2\pi}.\label{25}
\end{equation}

The infinitesimal change in the distribution function in the strong 
magnetic field is thus obtained from the RBTE 
in the relaxation-time approximation
\begin{equation}
p^{0}\frac{\partial f_{i,B}}{\partial t}+
p^{3}\frac{\partial f_{i,B}}{\partial x^3}
+q_iF^{03}p_3\frac{\partial f_{i,B}}{\partial 
p^0}+q_iF^{30}p_0\frac{\partial {f_{i,B}}}
{\partial {p_3}}=-\frac{p^0}{\tau^B}\delta 
f_{i,B},\label{26}
\end{equation}
where $\tau_i$ denotes the relaxation-time 
for quarks in the presence of strong magnetic field, which, in 
the Lowest Landau Level (LLL) approximation 
is given by\cite{Hattori:PRD95'2017}:
\begin{equation}
\tau_i (T,B)=\frac{w_i\left(e^{\beta \omega_i}
-1\right)}{\alpha_s \left(\Lambda^2, eB\right) C_2m_i^2\left(e^{\beta 
\omega_i}+1\right)}\left[\frac{1}{\int dp'^3 
\frac{1}{w'_i\left(e^{\beta \omega'_i}+1
\right)}}\right],\label{27}
\end{equation}
which has been evaluated for massless quarks. However, it has been shown in 
Ref.\cite{Berrehrah:PRC89'2014} that the effect of finite quark mass in the 
evaluation of scattering cross sections is very small, and hence, the 
relaxation-time is largely unaffected. $C_2=4/3$ is the 
Casimir factor. We use a one loop running coupling constant $\alpha_s (\Lambda^2,eB)$, 
which runs with both the magnetic field and temperature. In the strong magnetic field (SMF) regime, 
its form is given by:\cite{Ayala:PRD98'2018}
\begin{equation}
 \alpha_s(\Lambda^2,|eB|)=\frac{\alpha_s(\Lambda^2)}{1+b_1\alpha_s(\Lambda^2)\,\mbox{ln}\left(\frac{\Lambda^2}{\Lambda^2+|eB|}\right)}.\label{coupling}
\end{equation}
\vspace{4mm}
where, $\alpha_s(\Lambda^2)$ is the one-
	loop running coupling in the absence of a magnetic field.
	$$\alpha_s(\Lambda^2)=\frac{1}{b_1\,\mbox{ln}\left(\frac{\Lambda^2}{\Lambda_{QCD}^2}\right)},$$
	where $b_1=(11N_c-2N_f)/12\pi$ and $\Lambda_{QCD}\sim 0.2$ GeV. The renormalisation scale is chosen 
	to be $\Lambda=2\pi\sqrt{T^2+\frac{\mu^2}{\pi^2T^2}}$. Thus, via the strong coupling, 
	the relaxation time acquires an implicit dependence on the chemical 
potential.

Now, the infinitesimal change for quark and anti-quark distribution functions
can be obtained from the RBTE (eq.\ref{26}) in SMF regime as
\begin{align}
\delta f_{i,B}&=-\frac{\tau_i^B}{p^0}\frac
{p_z f_{i,B}(1-f_{i,B})}{T}\left[\frac{
\omega_i-\mu_i}{T}(\vec{\nabla}T)_z-2q_
iE_z\right]\label{29}\\[0.8em]
\bar{\delta f_{i,B}}&=-\frac{\tau_i^B}{p^0}
\frac{p_z \bar{f}_{i,B}(1-\bar{f}_{i,B})}{T}
\bigg[\frac{\omega_i+\mu_i}{T}(\vec{\nabla}T)_z
+2q_iE_z\bigg],\label{30}
\end{align}
which gives the induced current density from \eqref{24} for a single 
species,
\begin{multline}
J_{z}=(\vec{\nabla}T)_{z}\left[\frac{qg|qB|}{T(2\pi)^2}\int\frac{dp_z}{\omega^2}p_z^2(\tau^B)\left\{\bar{f}(1-\bar{f})\frac{\omega+\mu}{T}-f(1-f)\frac{\omega-\mu}
{T}\right\}\right]\\-2qE_z\left[\frac{qg|qB|}{T(2\pi)^2}\int \frac{dp_z}{\omega^2}p_z^2(\tau^B)\left\{\bar{f}(1-\bar{f})+f(1-f)\right\}\right],\label{31}
\end{multline}
where, $g$ is the degeneracy factor.
Defining the following integrals
\begin{eqnarray}
H_1&=&\int \frac{dp_z} {w_i^2}\,\tau_{B}^i\,p_z^2\left\{-\bar{f}
(1-\bar{f})(\omega+\mu)+f(1-f)(\omega-\mu)
\right\},\\
H_2&=&\int \frac{dp_z}{w_i^2}\,\tau_{B}^i\
,p_z^2\left\{\bar{f}(1-\bar{f})+f(1-f)
\right\},
\end{eqnarray}
the  current density (3rd-component) from eq.\eqref{31} can be recast in the form
\begin{equation}
J_{z}=(\vec{\nabla}T)_{z}\,\frac{qg|qB|}{T^2(2\pi)^2}H_1-E_z\,\frac{qg|qB|}{T(2\pi)^2}2qH_{2}\label{32}
\end{equation}

As earlier, the induced electric field due to the temperature-gradient is
obtained by setting $J_{z}=0$,
\begin{equation}
E_z=\frac{1}{2Tq}\frac{H_{1}}{H_{2}}(\vec{\nabla}T)_z.
\end{equation}
The proportionally constant gives the Seebeck coefficient:
\begin{equation}
S=\frac{1}{2Tq}\frac{H_{1}}{H_{2}}\label{34}
\end{equation}
\vspace{5mm}
\begin{figure}[H]
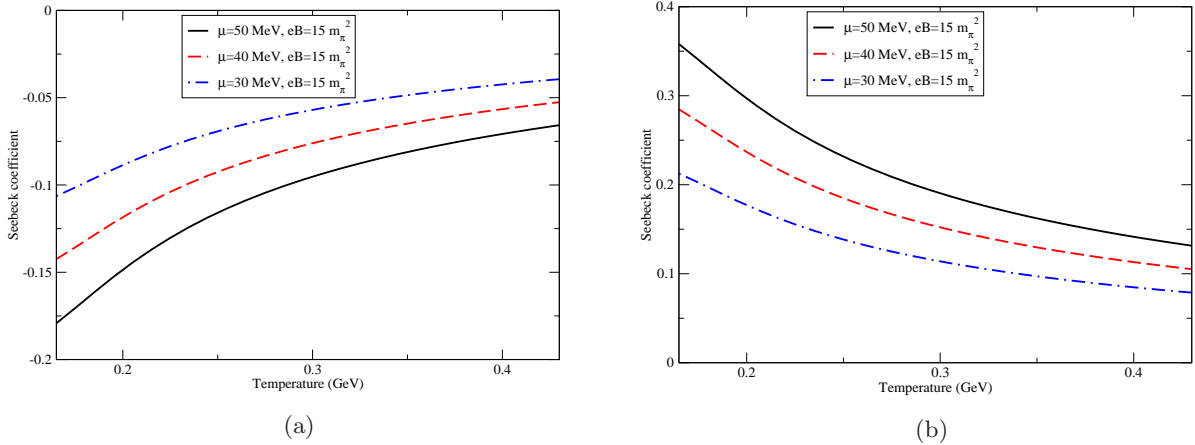

\begin{subfigure}{0.48\textwidth}
\includegraphics[width=0.95\textwidth]{uquark.eps}
\caption{}\label{fig6a}
\end{subfigure}
\hspace*{\fill}
\begin{subfigure}{0.48\textwidth}
\includegraphics[width=0.95\textwidth]{dquark.eps}
\caption{}\label{fig6b}
\end{subfigure}
	\caption{Variation of Seebeck coefficient of 
$u$ (left) and $d$ (right) quarks with 
temperature for a fixed chemical potential and 
magnetic field.}
\label{fig6}
\end{figure}
As can be seen from Fig.(\ref{fig6}), the variation of the individual Seebeck coefficients (magnitudes) of the $u$ and $d$ 
quarks with temperature and chemical potential shows the same trend as in the earlier case. However, the sign of the 
Seebeck coefficient in this case is negative for $u$ quark and positive for $d$ quark. This is opposite to what was 
encountered earlier. This is because the $H_1$ integrals for both $u$ and $d$ quarks turn out to be 
negative in this case (Fig.(\ref{fig8})) and the $H_2$ integrals positive (Fig.(\ref{fig9})). Considered along with the 
dependence on the particle charge, this explains the sign of the Seebeck coefficient for $u$ and $d$ quarks.  
The Seebeck coefficients for the $u$ and $d$ quarks in this case turn out to be about 1 
order of magnitude smaller, compared to their $B=0$ counterparts.    
\vspace{6mm}
 \begin{figure}[H]
	\centering
	\includegraphics[width=9.5cm,height=9.2cm,keepaspectratio]{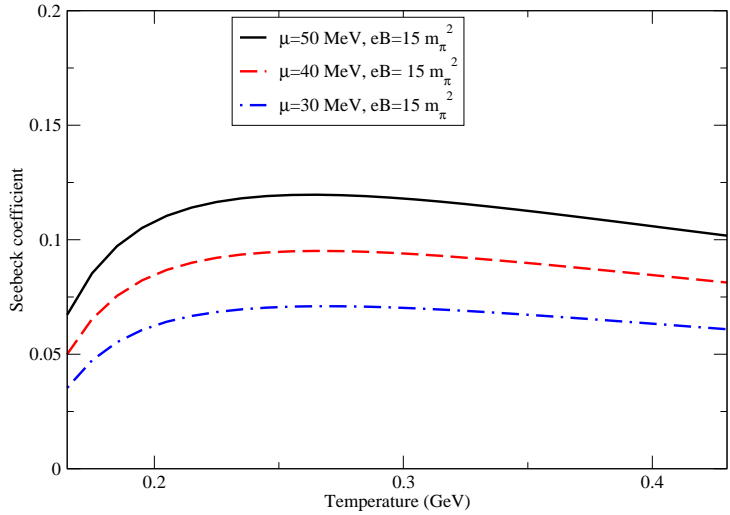}
	\caption{Variation of $s$ quark Seebeck 
coefficient with temperature for a fixed 
magnetic field and 
different fixed values of chemical potential.}
\label{fig7}
\end{figure}
The sign of the $s$ quark Seebeck coefficient is again opposite to that of the $B=0$ 
case owing to the $H_1$ integral for $s$ quark being negative. The magnitude of the Seebeck 
coefficient rises with temperature upto about $T=270$ MeV and starts decreasing therefrom. 
Although the ratio $H_1/H_2$ is an increasing function of $T$ for the entire temperature range, 
it does not increase fast enough after $T= 270$ MeV to compensate for the rising temperature (in the denominator of $S$; eq.\eqref{34}).

It should be noted that contrary to the
case of pure thermal medium, the relaxation 
time here is momentum dependent. As such, it 
cannot be taken out of the momentum integrations 
($H_{1},H_{2}$) and hence does not cancel 
out in $S$. 
\vspace{4mm}
\begin{figure}[H]
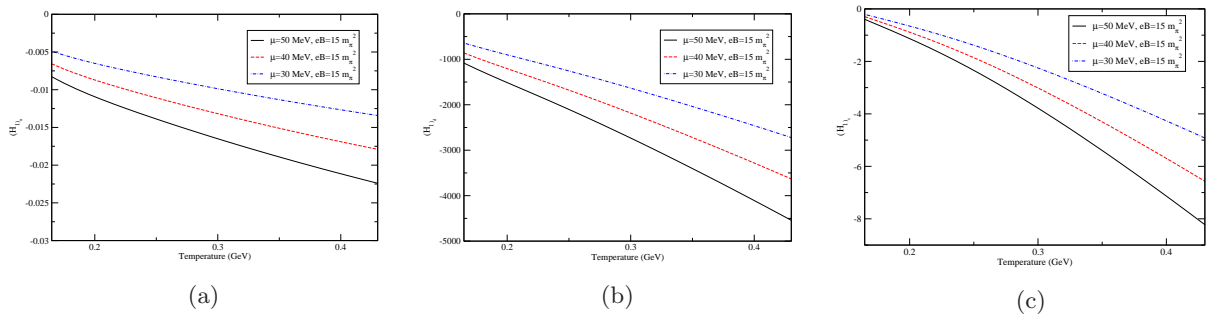

\begin{subfigure}{0.32\textwidth}
\includegraphics[width=0.95\textwidth]{H1unqpm.eps}
\caption{}\label{fig8a}
\end{subfigure}
\hspace*{\fill}
\begin{subfigure}{0.32\textwidth}
\includegraphics[width=0.95\textwidth]{H1dnqpm.eps}
\caption{}\label{fig8b}
\end{subfigure}
\hspace*{\fill}
\begin{subfigure}{0.32\textwidth}
\includegraphics[width=0.95\textwidth]{H1snqpm.eps}
\caption{}\label{fig8c}
\end{subfigure}
	\caption{Variation of $H_1$ integrals for $u$ (left), $d$ (middle) and $s$ (right) quarks with temperature}
	\label{fig8}
	\end{figure}
	
\begin{figure}[H]
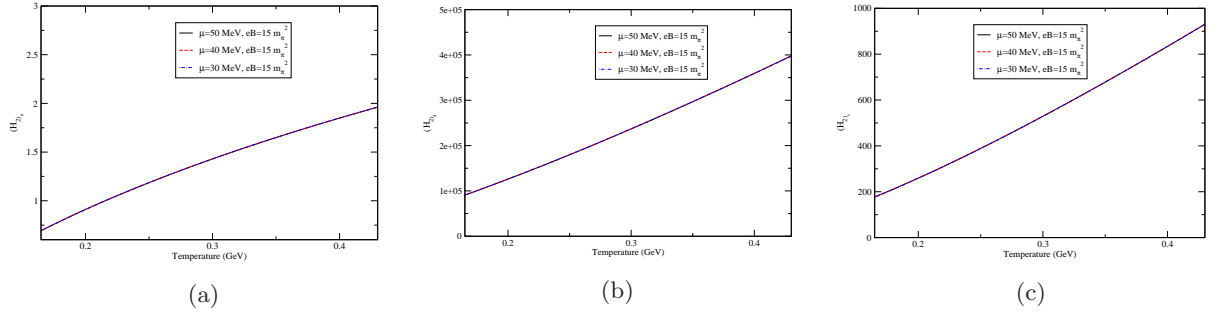

\begin{subfigure}{0.32\textwidth}
\includegraphics[width=0.95\textwidth]{H2unqpm.eps}
\caption{}\label{fig9a}
\end{subfigure}
\hspace*{\fill}
\begin{subfigure}{0.32\textwidth}
\includegraphics[width=0.95\textwidth]{H2dnqpm.eps}
\caption{}\label{fig9b}
\end{subfigure}
\hspace*{\fill}
\begin{subfigure}{0.32\textwidth}
\includegraphics[width=0.95\textwidth]{H2snqpm.eps}
\caption{}\label{fig9c}
\end{subfigure}
	\caption{Variation of $H_2$ integrals for $u$ (left), $d$ (middle) and $s$ (right) quarks with temperature.}
	\label{fig9}
	\end{figure}

Now we generalize our formalism to a medium consisting of multiple species, 
therefore, the total current is given by the sum of currents due to individual species:
\begin{eqnarray}
J_z &=&J_z^1+J_z^2+J_z^3+ \cdots \nonumber\\
&=&(\vec{\nabla}T)_{z}\left\{\frac{q_1g_1|q_1B|}{T^2(2\pi)^2}(H_{1})_1+\frac{q_2g_2|q_2B|}{T^2(2\pi)^2}(H_{1})_2+\frac{q_3g_3|q_3B|}{T^2(2\pi)^2}(H_{1})_3+
\cdots \right\} \nonumber\\
&-&E_z\left\{\frac{q_1g_1|q_1B|}{T(2\pi)^2}2q_1(H_{2})_1+\frac{q_2g_2|q_2B|}{T(2\pi)^2}2q_2(H_{2})_2+\frac{q_3g_3|q_3B|}{T(2\pi)^2}2q_3(H_{2})_3+ \cdots
\right\}\label{35}
\end{eqnarray}
Again, the Seebeck coefficient of the medium in a strong magnetic field
is obtained by setting $J_z=0$,
\begin{equation}
S=\frac{1}{2T}\frac{\sum_i q_i|q_iB|(H_1)_i}
{\sum_i q_i^2|q_iB|(H_2)_i},\label{37}
\end{equation}
which could be further expressed in terms of the weighted average of 
individual Seebeck coefficients.
\begin{equation}
S=\frac{\sum_iS_i|q_i|^3(H_2)_i}{\sum_i|q_i|
^3(H_2)_i}.\label{38}
\end{equation}
Thus, unlike the Seebeck coefficient in the absence of magnetic field (eq.\ref{13}), 
both the individual as well as total Seebeck coefficient of the medium 
depend on the relaxation-time in the presence of a strong 
magnetic field. 
\begin{figure}[H]
	\centering
	\includegraphics[width=9.5cm,height=9.2cm,keepaspectratio]{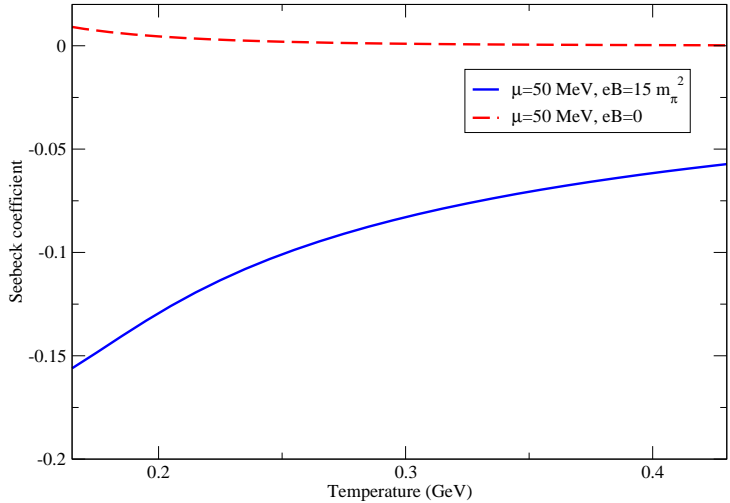}
	\caption{Variation of Seebeck coefficient 
	of the medium with temperature for a 
	fixed chemical potential in the absence 
	and presence of a magnetic field.}
\label{fig10}
	\end{figure}
We can now visualize the sole effect of strong magnetic field on the (total) 
Seebeck coefficient from the comparison of $B \neq 0$ and $B=0$ results
in Fig.(\ref{fig10}). Unlike the $B=0$ (red line) case, the total Seebeck 
coefficient in strong $B$ (black line) becomes negative, 
indicating that the induced electric field is  opposite to the direction 
of temperature-gradient. 
However, the magnitude of total Seebeck coefficient decreases with the
temperature and increases with the chemical potential, 
{\em much like}, the $B=0$ case. However, the strong magnetic field 
enhances the magnitude of $S$
by one order of magnitude, compared to the $B=0$ case. 

\section{Seebeck effect of hot partonic medium in a quasiparticle model}
Quasiparticle description of quarks and gluons in a thermal QCD medium 
in general, introduces a thermal mass, apart from their current masses in QCD 
Lagrangian. These masses are generated due to the 
interaction of a given parton with other partons in the medium, therefore, 
quasiparticle description in turn describes the 
collective properties of the medium. 
However, in the presence of strong magnetic field in the thermal
QCD medium, different flavors 
acquire masses differently due to their different electric charges.
Different versions of quasiparticle 
description exist in the literature based on different effective theories, 
{\em such as} Nambu-Jona-Lasinio (NJL) model and its extension PNJL model~\cite{Fukushima:PLB591'2004,Ghosh:PRD73'2006,Abuki:PLB676'2009}, 
Gribov-Zwanziger quantization \cite{Su:PRL114'2015,Florkowski:PRC94'2016}, thermodynamically consistent quasiparticle model \cite{Bannur:JHEP0709'2007}, 
etc. However, our description relies on perturbative thermal QCD, where the 
medium generated masses for quarks and gluons are obtained from the poles 
of dressed propagators calculated by the respective self-energies at 
finite temperature and/or strong magnetic field.

\subsection{Seebeck coefficient in the absence of magnetic field}
In the quasiparticle description of quarks and gluons
in a thermal medium with $3$ flavours, all flavours
(with current/vacuum masses, $m_i << T$) 
acquire the same thermal mass~\cite{Braaten:PRD45'1992,Peshier:PRD66'2002}
\begin{equation}
m_T^2=\frac{g^2(T)T^2}{6},\label{Gluon mass}
\end{equation}
which is, however, modified in the presence of a finite chemical 
potential~\cite{Kakade:PRC92'2015}
\begin{equation}\label{Quark mass}
m_{T,\mu}^2=\frac{g^2(T)T^2}{6}\left(1+\frac{\mu^2}{\pi^2T^2}\right),
\end{equation}
where $g$ is the running coupling constant already mentioned in eq.\eqref{100}.\\ 

We take the quasiparticle mass (squared) of $i^{\mbox{th}}$ 
flavor in a pure thermal medium to be~\cite{Bannur:JHEP0709'2007}:
\begin{equation}
m_{iT}^{\prime~2}=m_{i}^2+\sqrt{2}\,m_{i}\,m_T+m_T^2,\label{em}
\end{equation}
where $m_{i}$ is the current quark mass of the $i^{\mbox{th}}$ 
flavour. So the dispersion relation for the $i^{\mbox{th}}$ flavour 
takes the form
\begin{equation}
\omega_i^2(p)=\vec{p_i}^{\,\,2}+m_{i}^2+\sqrt{2}\,m_{i}\,m_T+m_T^2.\label{39}
\end{equation}
Using this expression of $\omega_i(p)$ in the quark distribution functions
as well as in the integrals $I_1$ and $I_2$ (eq.\ref{two}), we
proceed in a similar fashion and evaluate the individual Seebeck coefficients 
for the $u$, $d$ and $s$ quarks in quasiparticle description from eq.\eqref{13}
\vspace{7mm}
\begin{figure}[H]
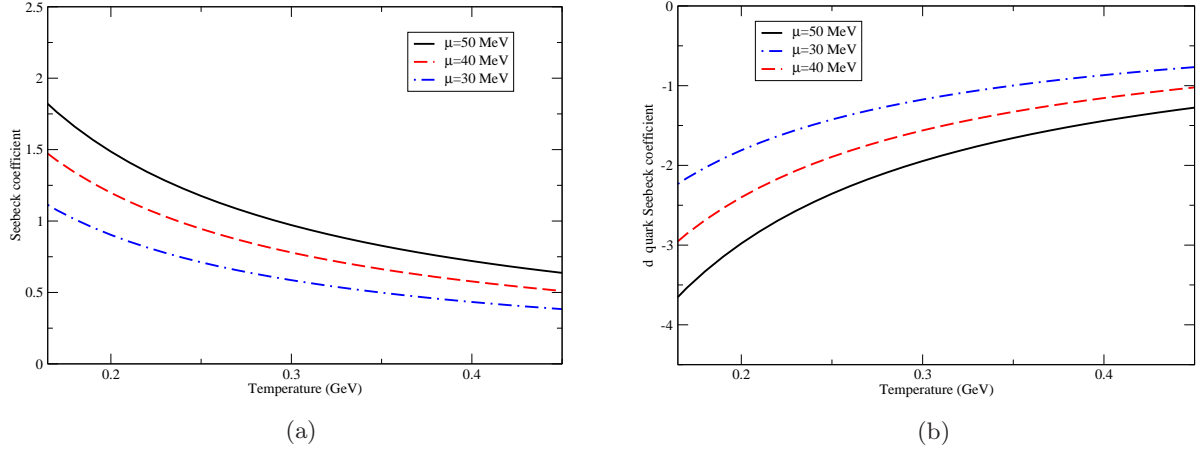

\begin{subfigure}{0.48\textwidth}
\includegraphics[width=0.95\textwidth]{newu.eps}
\caption{}\label{fig11a}
\end{subfigure}
\hspace*{\fill}
\begin{subfigure}{0.48\textwidth}
\includegraphics[width=0.95\textwidth]{newd.eps}
\caption{}\label{fig11b}
\end{subfigure}
	\caption{Variation of Seebeck coefficient of 
$u$ (left) and $d$ (right) quarks with 
temperature for different fixed values of 
chemical potentials.}
\label{fig11}
\end{figure}
As can be seen from Figs. \ref{fig11a} and \ref{fig11b}, the Seebeck coefficients of $u$ 
and $d$ quarks show a trend similar to their current quark mass counterparts 
in the absence of magnetic field (Figs. \ref{fig1a} \& \ref{fig1b}) and their
magnitudes decrease with temperature 
and increase with chemical potential. The $I_1$ and $I_2$ integrals
for both quarks are also found to be positive and as such, the sign of 
the coefficient is again determined by the electric charge of the 
particle. The change due to the quasiparticle description adopted here, 
is a slight increase in the magnitudes of the Seebeck coefficients for 
both quarks. 
\vspace{9mm}
\begin{figure}[H]
	\centering
	\includegraphics[width=9.5cm,height=9.2cm,keepaspectratio]{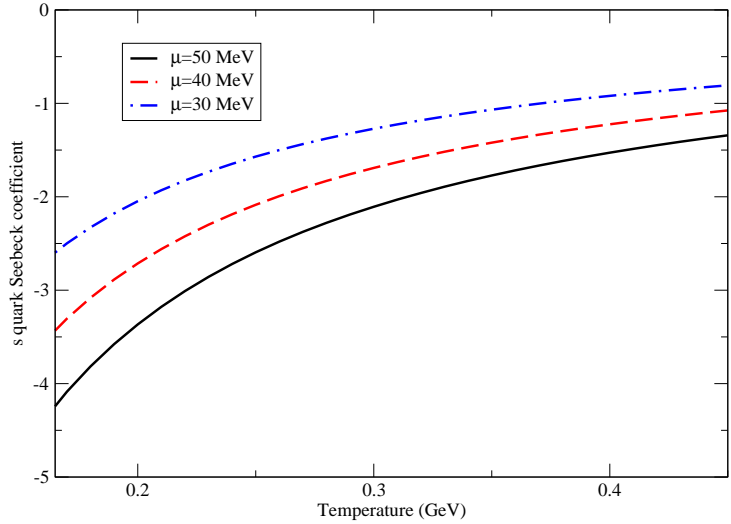}
	\caption{Variation of $s$ quark Seebeck 
coefficient with temperature for different 
fixed values of chemical potentials.}
\label{fig12}
\end{figure}
Similar to the current quark mass case (in Fig. \ref{Fig2}), the coefficient 
for $s$ quark decreases in magnitude with increasing temperature 
(Fig. \ref{fig12}). Owing to its negative electric charge and the positive 
value of $I_1$, $I_2$ integrals, the sign of the coefficient is negative.
Thus, the overall behaviour of individual Seebeck coefficients in the 
quasiparticle description is similar to that of the current mass description 
with slightly enhanced magnitudes. Numerically the average percentage 
increase for the $u$, 
$d$ and $s$ quarks are around 19.52\%, 19.70\% and 24.38\%, respectively.\\

Similarly, the total Seebeck coefficient of the medium in quasiparticle
description
\vspace{9mm}
\begin{figure}[H]
	\centering
	\includegraphics[width=9.5cm,height=9.2cm,keepaspectratio]{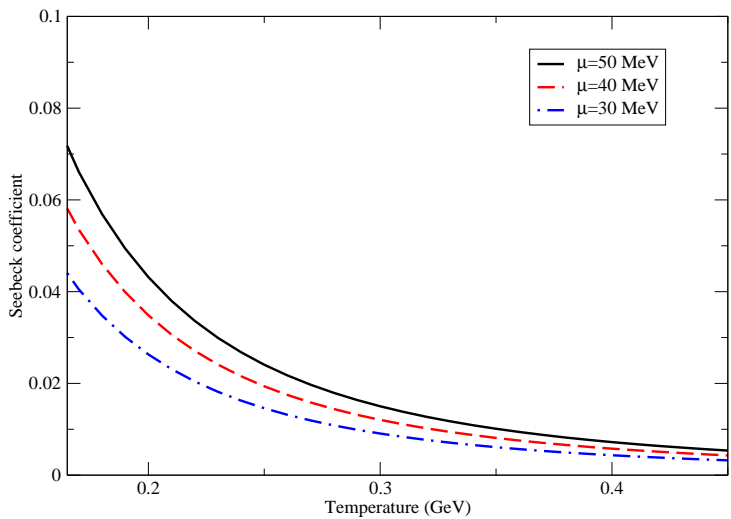}
	\caption{Variation of Seebeck coefficient of the medium with temperature for different 
fixed values of chemical potential.}
\label{fig16}
\end{figure}
(in Fig. \ref{fig16}) is found to have a small positive value, which
decreases with the temperature and increases with the chemical 
potential as earlier, but with a significantly 
elevated magnitude in comparison with the current quark mass case (in Fig \ref{fig5}). 

\subsection{Seebeck coefficient in the presence of strong magnetic field}
In the presence of magnetic field, only quarks are affected while the gluons 
are not directly influenced. As a result, only the quark-loop 
of the gluon self-energy will be affected and the gluon-loops remains altered.
Furthermore, only quarks contribute to the thermoelectric effect, and hence, we
proceed to calculate the thermal quark mass in the presence 
of a strong magnetic field, which can be obtained from the pole 
($p_0=0, \mathbf{p}\rightarrow 0$ limit) of the full
propagator. 

As we know, the full quark propagator can be obtained 
self-consistently from the Schwinger-Dyson equation (assuming massless
flavours, which is assumed to be true at least for light flavours), 
\be\label{S.D.E.}
S^{-1}(p_\parallel)=\gamma^\mu p_{\parallel\mu}-\Sigma(p_\parallel)
~,\ee
where $\Sigma (p_\parallel)$ is the quark self-energy at 
finite temperature in the presence of strong magnetic 
field. We can evaluate it up to one-loop
\begin{eqnarray}\label{Q.S.E.}
\Sigma(p)=-\frac{4}{3} g_s^{2}i\int{\frac{d^4k}{(2\pi)^4}}
\left[\gamma_\mu {S(k)}\gamma_\nu {D^{\mu \nu} (p-k)}\right]
,\end{eqnarray}
where $4/3$ denotes the Casimir factor and $g_s=\sqrt{4\pi\alpha_s}$ represents the running coupling with the $\alpha_s$ already defined in Eq.\eqref{coupling}. 
$D^{\mu \nu}(p-k)$ is the gluon propagator, which is not affected 
by the magnetic field, so its form is given by 
\be
\label{g. propagator}
D^{\mu \nu} (p - k)=\frac{ig^{\mu \nu}}{{(p-k)}^2}
~.\ee
However, the quark propagator, $S(K)$ in the strong magnetic 
field limit, is affected and is obtained by the Schwinger proper-time 
method at the lowest Landau level (n=0) in momentum space,
\be\label{q. propagator}
S(k)=ie^{-\frac{k^2_\perp}{|q_iB|}}\frac{\left(\gamma^0 k_0-\gamma^3 k_z+m_i\right)}{k^2_\parallel-m^2_i}\left(1
-\gamma^0\gamma^3\gamma^5\right)\label{exp},
\ee
where the 4-vectors are defined as $k_{\perp}\equiv(0,k_x,k_y,0), ~~ 
k_{\parallel}\equiv(k_0,0,0,k_z)$.

Next we obtain the form of quark and gluon propagators
at finite temperature in the imaginary-time formalism and
subsequently replace the energy integral ($\int \frac{dp_0}{2\pi}$)
by Matsubara frequency sum. However, in a strong
magnetic field along $z$-direction, the transverse component
of the momentum becomes vanishingly small ($k_\perp \approx 0$), so
the exponential factor in eq.\eqref{exp} becomes unity and the
integration over the transverse component of the momentum becomes
$|q_fB|$. Thus, the quark self-energy in eq.\eqref{Q.S.E.} at finite
temperature in the SMF limit will be of the form 
\begin{eqnarray}\label{Q.S.E.(1)}
\nonumber\Sigma(p_\parallel) &=& \frac{2g_s^2}{3\pi^2}|q_iB|T\sum_n\int dk_z\frac{\left[\left(1+\gamma^0\gamma^3\gamma^5\right)\left(\gamma^0k_0
-\gamma^3k_z\right)-2m_i\right]}{\left[k_0^2-\omega^2_k\right]\left[(p_0-k_0)^2-\omega_{pk}^2\right]} \\ &=& \frac{2g_s^2|q_iB|}{3\pi^2}\int dk_z\left[(\gamma^0+\gamma^3\gamma^5)L^1-(\gamma^3+\gamma^0\gamma^5)k_zL^2\right]
,\end{eqnarray}
where $\omega^2_k=k_z^2+m_i^2$,\, $\omega_{pk}^2=(p_z-k_z)^2$ and
$L^1$ and $L^2$ are the two frequency sums, which are given by
\be
&&L^1=T\sum_nk_0~\frac{1}{\left[k_0^2-\omega_k^2\right]}\frac{1}{\left[(p_0-k_0)^2-\omega_{pk}^2\right]} ~, \\ &&L^2=T\sum_n\frac{1}{\left[k_0^2-\omega_k^2\right]}\frac{1}{\left[(p_0-k_0)^2-\omega_{pk}^2\right]}
~.\ee

We first do the frequency sums \cite{Kapusta:Book'2006,Bellac:BOOK'1996} 
and then integrate the momentum $k_z$ to obtain the simplified form of 
quark self-energy eq.\eqref{Q.S.E.(1)}~\cite{Rath:EPJA55'2019} as
\begin{eqnarray}\label{Q.S.E.(3)}
\Sigma(p_\parallel)=\frac{g_s^2|q_iB|}{3\pi^2}\left[\frac{\pi T}{2m_i}-\ln(2)\right]\left[\frac{\gamma^0p_0}{p_\parallel^2}+\frac{\gamma^3p_z}{p_\parallel^2}+\frac{\gamma^0\gamma^5p_z}{p_\parallel^2}+\frac{\gamma^3\gamma^5p_0}{p_\parallel^2}\right]
.\end{eqnarray}

To solve the Schwinger-Dyson equation eq.\eqref{S.D.E.}, one needs to first express the self-energy at finite temperature in 
magnetic field in a covariant form~\cite{Ayala:PRD91'2015,Karmakar:1902.02607},
\begin{equation}\label{1general q.s.e.}
\Sigma(p_\parallel)=A {(\rm p_0,\bf p)} \gamma^\mu u_\mu+ B {(\rm p_0,\bf p)} 
\gamma^\mu b_\mu + C {(\rm p_0, \bf p)} \gamma^5\gamma^\mu u_\mu +D
{(\rm p_0, \bf p)} \gamma^5\gamma^\mu b_\mu
~,\end{equation}
where $u^\mu$ (1,0,0,0) and $b^\mu$ (0,0,0,-1) denote the preferred directions 
of heat bath and magnetic field, respectively and these vectors mimic the 
breaking of Lorentz and rotational invariances, respectively. 
The form factors, $A$, $B$, $C$ and $D$ are computed in strong $B$ with LLL 
approximation as
\begin{eqnarray}
&&A=\frac{g_s^2|q_iB|}{3\pi^2}\left[\frac{\pi T}{2m_i}-\ln(2)\right]\frac{p_0}{p_\parallel^2} ~, \\ 
&&B=\frac{g_s^2|q_iB|}{3\pi^2}\left[\frac{\pi T}{2m_i}-\ln(2)\right]\frac{p_z}{p_\parallel^2} ~, \\ 
&&C=-\frac{g_s^2|q_iB|}{3\pi^2}\left[\frac{\pi T}{2m_i}-\ln(2)\right]\frac{p_z}{p_\parallel^2} ~, \\ 
&&D=-\frac{g_s^2|q_iB|}{3\pi^2}\left[\frac{\pi T}{2m_i}-\ln(2)\right]\frac{p_0}{p_\parallel^2}
~.\end{eqnarray}
Then the self-energy \eqref{1general q.s.e.} can be expressed in terms 
of chiral projection operators ($P_R$ and $P_L$) as
\begin{equation}\label{projection1}
\Sigma(p_\parallel)=P_R\left[(A-B)\gamma^\mu u_\mu+(B-A)\gamma^\mu b_\mu
\right]P_L+P_L\left[(A+B)\gamma^\mu u_\mu+(B+A)\gamma^\mu b_\mu\right]P_R
~.\end{equation}

Hence, the Schwinger-Dyson equation is able to express the inverse of the full 
propagator in terms of $P_L$ and $P_R$,
\be\label{prop}
S^{-1}(p_\parallel)=P_R\gamma^\mu X_\mu P_L+P_L\gamma^\mu Y_\mu P_R
~,\ee
where
\begin{eqnarray}
&&\gamma^\mu X_\mu=\gamma^\mu p_{\parallel\mu}-(A-B)\gamma^\mu u_\mu-(B-A)\gamma^\mu b_\mu ~, \\
&&\gamma^\mu Y_\mu=\gamma^\mu p_{\parallel\mu}-(A+B)\gamma^\mu u_\mu-(B+A)\gamma^\mu b_\mu
~.\end{eqnarray}
Thus, the effective propagator is finally obtained by inverting eq. \eqref{prop}
\be
S(p_\parallel)=\frac{1}{2}\left[P_R\frac{\gamma^\mu Y_\mu}{Y^2/2}P_L+
P_L\frac{\gamma^\mu X_\mu}{X^2/2}P_R\right]
,\ee
where
\begin{eqnarray}
&&\frac{X^2}{2}=X_1^2=\frac{1}{2}\left[p_0-(A-B)\right]^2-\frac{1}{2}\left[p_z+(B-A)\right]^2 ~, \\
&&\frac{Y^2}{2}=Y_1^2=\frac{1}{2}\left[p_0-(A+B)\right]^2-\frac{1}{2}\left[p_z+(B+A)\right]^2
~.\end{eqnarray}
Thus, the thermal mass (squared) for $i^{\mbox{th}}$ flavor 
at finite temperature and strong
magnetic field is finally obtained by taking the $p_0=0, p_z\rightarrow 0$
limit in either $X_1^2$ or $Y_1^2$ (because both of them are equal),
\begin{eqnarray}\label{Mass}
m_{iT,B}^2 =\frac{g_s^2|q_iB|}{3\pi^2}\left[\frac{\pi T}{2m_i}-\ln(2)\right]
,\end{eqnarray}
which depends both on temperature and magnetic field. The quark distribution functions with medium generated masses in the 
absence and presence of magnetic field therefore 
manifest the interactions present in the respective 
medium in terms of modified occupation probabilities in the phase space and
thus affect the Seebeck coefficients.
The quasiparticle (or effective) mass of $i^{\mbox{th}}$ 
quark flavor is generalized in finite temperature and strong
magnetic field into 
\begin{equation}
m_{i~T,B}^{\prime~2}=m_{i}^2+\sqrt{2}\,m_{i}\,m_{iT,B}+m_{iT,B}^2.
\end{equation}

\begin{figure}[H]
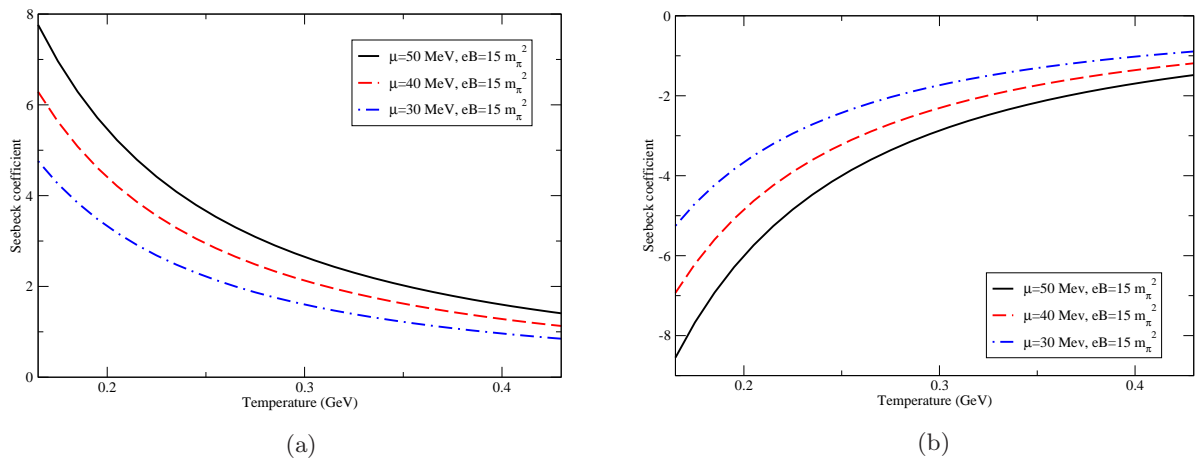

\begin{subfigure}{0.48\textwidth}
\includegraphics[width=0.95\textwidth]{mnsu.eps}
\caption{}\label{fig19a}
\end{subfigure}
\hspace*{\fill}
\begin{subfigure}{0.48\textwidth}
\includegraphics[width=0.95\textwidth]{mnsd.eps}
\caption{}\label{fig19b}
\end{subfigure}
	\caption{Variation of Seebeck coefficient of 
$u$ (left) and $d$ (right) quarks with 
temperature for a fixed chemical potential and 
magnetic field.}
\label{fig19}
\end{figure}

Now, using the above quasiparticle mass in the distribution function 
and proceeding identically, we compute the individual Seebeck coefficients 
from eq.\eqref{34} as a function of temperature, which is shown in Fig. \ref{fig19}.
\vspace{8mm}
We find that the magnitudes of both $u$ and $d$ quark Seebeck coefficients 
decrease with the temperature and increase with the chemical potential. In
quasiparticle description, the $H_1$ integrals for both $u$ and $d$ quarks 
becomes positive, {\em unlike} the case with current quark masses
in strong $B$ (Fig. \ref{fig6}), so the sign of the individual Seebeck 
coefficient is decided only by the electric charge of 
quarks, similar to the $B=0$ case. 
The magnitudes of the Seebeck coefficients are found to increase 
by two-order of magnitude over their 
current quark mass case counterparts (seen in Fig. \ref{fig6}), which could thus
be attributed to the quasiparticle description.  
\vspace{8mm}
\begin{figure}[H]
\centering
	\includegraphics[width=9.5cm,height=9.2cm,keepaspectratio]{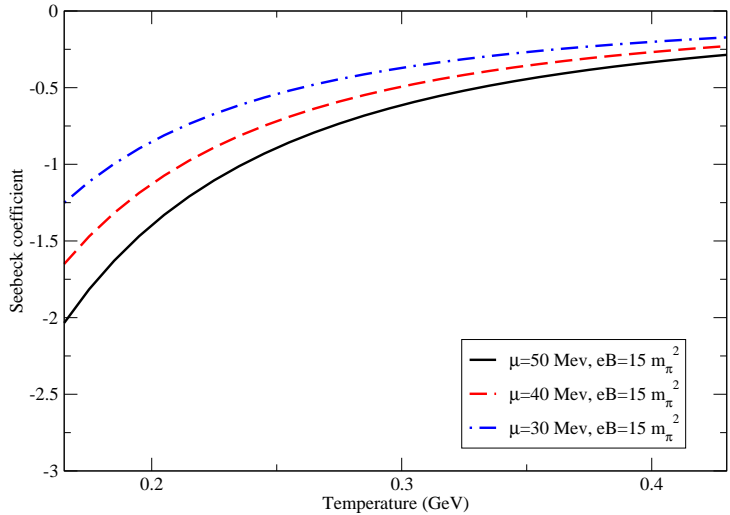}
	\caption{Variation of $s$ quark Seebeck 
coefficient with temperature for a fixed 
magnetic field and 
different fixed values of chemical potential.}
\label{fig20}
\end{figure}
The sign of the Seebeck coefficient for $s$-quark 
in quasiparticle description now becomes negative, opposite to the case of 
current quark mass description (Fig.\ref{fig7} ). Again, this is because the quasiparticle
description flips the signs of $H_1$ integral for the $s$ 
quark from negative (in Fig.\ref{fig8}) to positive,  so the deciding factor
for the sign of the coefficient is the sign of the electric charge of $s$ quark 
(which is negative). Furthermore, the variation of the 
magnitude of Seebeck coefficient with temperature in quasiparticle
description is quite different compared to the current quarks mass case (Fig.\ref{fig7}) and is
rather similar to that of $d$ quark (Fig. \ref{fig19b}) but with smaller magnitude. 

Comparison between the $B=0$ and $B\neq 0$ results within the quasiparticle description
reveals summarily that the percentage increase is more 
pronounced at lower temperatures. The average percentage increase over the entire temperature range is 467.61\% and 212.63\% 
for $u$ and $d$ quarks, respectively. However, the percentage increase for 
$s$ quark is -36.81\%, suggesting that the $s$ quark Seebeck 
coefficient in the presence of a strong $B$ is smaller in 
magnitude than its pure thermal ($B=0$) counterpart (Fig.\ref{fig12}). 
Hence, in the presence of strong $B$, the Seebeck effect depends strongly 
on the interactions among the constituents in the medium, encoded by the 
appropriate quasiparticle description. 
\vspace{2mm}
\begin{figure}[H]
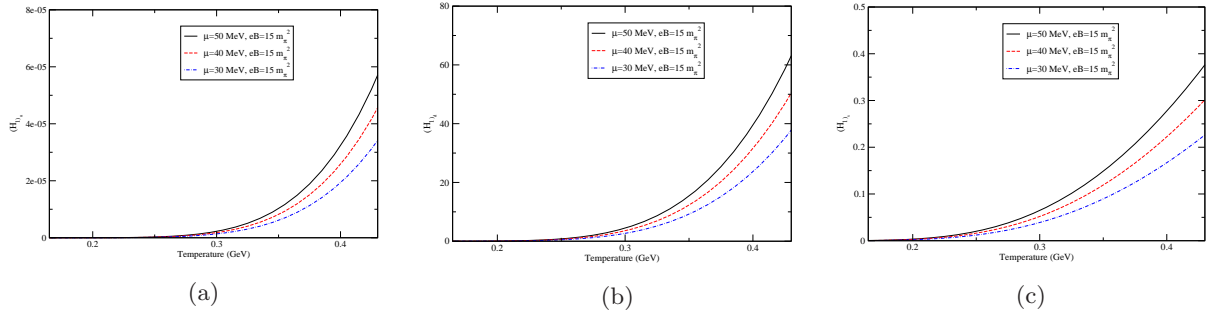

\begin{subfigure}{0.32\textwidth}
\includegraphics[width=0.95\textwidth]{H1uqpm.eps}
\caption{}\label{fig20a}
\end{subfigure}
\hspace*{\fill}
\begin{subfigure}{0.32\textwidth}
\includegraphics[width=0.95\textwidth]{H1dqpm.eps}
\caption{}\label{fig20b}
\end{subfigure}
\hspace*{\fill}
\begin{subfigure}{0.32\textwidth}
\includegraphics[width=0.95\textwidth]{H1sqpm.eps}
\caption{}\label{fig20c}
\end{subfigure}
\caption{Variation of $H_1$ integrals for $u$ (left), $d$ (middle) and $s$ (right) quarks with temperature for different 
fixed values of chemical potential.}\label{Fig20}
\end{figure}

\begin{figure}[H]
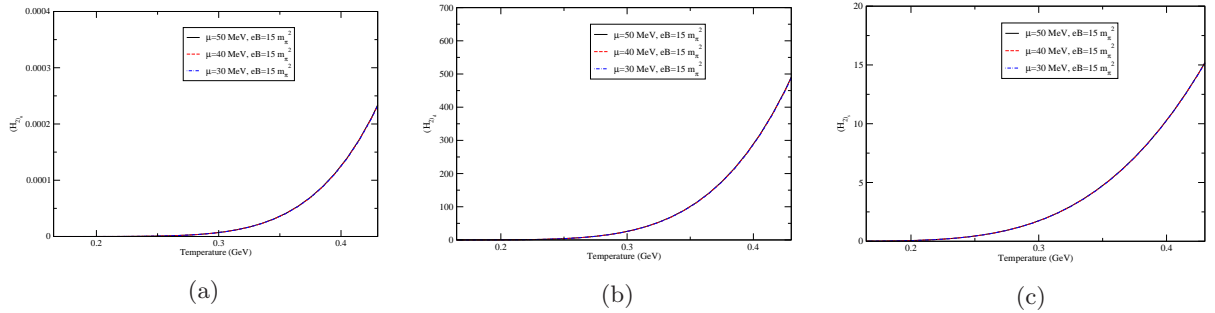

\begin{subfigure}{0.32\textwidth}
\includegraphics[width=0.95\textwidth]{H2uqpm.eps}
\caption{}\label{fig21a}
\end{subfigure}
\hspace*{\fill}
\begin{subfigure}{0.32\textwidth}
\includegraphics[width=0.95\textwidth]{H2dqpm.eps}
\caption{}\label{fig21b}
\end{subfigure}
\hspace*{\fill}
\begin{subfigure}{0.32\textwidth}
\includegraphics[width=0.95\textwidth]{H2sqpm.eps}
\caption{}\label{fig21c}
\end{subfigure}
\caption{Variation of $H_2$ integrals for $u$ (left), $d$ (middle) and $s$ (right) quarks with temperature for different 
fixed values of chemical potential.}\label{Fig21}
\end{figure}

\begin{figure}[H]
	\centering
	\includegraphics[width=9.5cm,height=9.2cm,keepaspectratio]{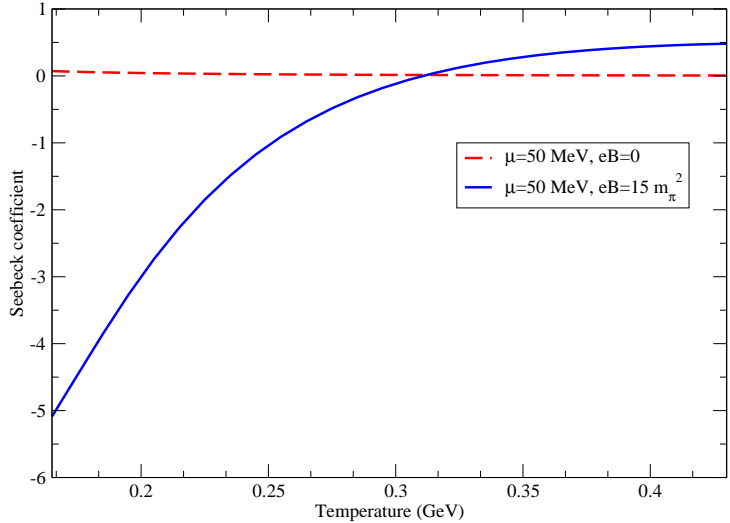}
	\caption{Variation of Seebeck coefficient 
	of the medium with temperature for a 
	fixed chemical potential in the absence 
	and presence of a magnetic field.}
\label{fig23}
\end{figure}

Once the individual Seebeck coefficients of $u$, $d$ and $s$ quarks 
have been evaluated in quasiparticle description, we compute
the weighted average of the above individual coefficients to obtain 
the (total) Seebeck coefficient of the medium as a function of temperature from eq.\eqref{38}. This is shown
in Fig. \ref{fig23}. To see the effects of magnetic field in quasiparticle
description, we have also displayed the same in the absence of magnetic
field (shown by black solid line in Fig. \ref{fig16}) in the same figure for better
visual effects.
\vspace{7mm}
The seebeck coefficient starts out negative and one order of magnitude larger than its immediate 
	counterpart in the absence of magnetic field. The magnitude decreases rapidly with increasing temperature, 
	eventually crosses the zero mark and continues to higher positive values thereafter. 
	Physically, the direction of the induced field is opposite to that of the temperature gradient to start with. 
	As the temperature rises, the strength of this field gets weaker. At a particular value of the temperature, 
	the individual seebeck coefficients have values so as to make the weighted average zero. For still higher 
	values of temperature, the weighted average becomes positive, indicating that the induced electric field is now along the temperature gradient.

\section{Summary and conclusions}
In this paper, we have investigated the thermoelectric phenomenon of 
Seebeck effect in a hot QCD medium in two descriptions: i) when the 
quarks are treated in QCD with their current masses and ii) when
the quarks are treated in quasiparticle model. The emergence of a strong 
magnetic 
field in the non-central events of the ultra-relativistic heavy ion 
collisions provides a further impetus to carry out the aforesaid 
investigations both in the absence and presence of a strong magnetic field, 
in order to isolate the effects of strong magnetic fields and interactions
present among partons. For this 
purpose, the Seebeck coefficients are calculated individually for the $u$, 
$d$ and $s$ quarks, which, in turn, give the Seebeck coefficient of the 
medium via a weighted average of the individual coefficients. Thus,
effectively, four different scenarios have been analysed:
\begin{enumerate}
	\item Current mass description with $B=0$. 
	\item Current mass description with $B\neq 0$.
	\item Quasiparticle description with $B=0$. 
	\item Quasiparticle description with $B\neq 0$.
\end{enumerate}

Comparison between the cases 1 and 3 is able to decipher 
the effect of the intereactions among the partons through the 
quasiparticle description on the
Seebeck effect in the absence of strong magnetic field, where
the magnitudes of Seebeck coefficient of individual species as well as 
that of the medium get(s) sightly 
enhanced with respect to the current mass 
description. The sign of the individual Seebeck 
coefficients is positive for positively charged particles ($u$ quark) and 
negative for negatively charged particles ($d$ and $s$ quarks) except for the 
case 2, where, owing to the $H_1$ integral becoming negative (for each quark), the situation is reversed.
Comparison between 
cases 2 and 4 brings out the sole effect of the quasiparticle description 
in the presence of a strong 
magnetic field, where it is seen that the magnitude of the coefficients are amplified. The sign of the 
Seebeck coefficient of the medium, is however same in both 
the cases. Lastly, the comparison between cases 3 and 4 brings 
forth the sole effect of strong constant 
magnetic field on the Seebeck effect in the quasiparticle description. The variation 
of individual and total Seebeck coefficients 
with temperature and chemical potential are 
found to show similar trends in both the cases but with enhanced magnitudes 
in the latter case.

The trend of overall decrease (increase) of the magnitude of Seebeck 
coefficient 
with the increase in temperature (chemical potential) is seen for all cases. 
However, in the quasiparticle description, the magnitude 
of the coefficient gets enhanced in the presence of strong magnetic field, 
so the inclusion of interactions among partons plays a crucial role in 
thermoelectric phenomenon in thermal QCD.

\section{Acknowledgements}
We are thankful to Sumit and Pushpa for their fruitful discussin
in the calculaton for the matrix element.
BKP is thankful to the Council of Scientific and
Industrial Research Grant No.03 (1407)/17/EMR-II),
Government of India for the financial assistance.

\appendix
\appendixpage
\addappheadtotoc
\begin{appendices}
\renewcommand \thesubsection{\Alph{section}.\arabic{subsection}}
\section{Matrix Element for gluon-gluon scattering in vacuum}
We will study the scattering in Figs. \ref{Fig1} and \ref{fgv} in the cm frame with the 
four momenta for incoming particles as $K_1 (k_1^0,\vec{k_1})$ and $K_2 (k_2^0,
\vec{k}_2)$, and that for
outgoing particles as $K_3 (k_3^0,\vec{k_3})$ and $K_4 (k_4^0,\vec{k_4})$.
Moreover, we assume that the trajectory of the incoming particles is
along the $z$-direction and the scattered particles lie in the $x$-$z$ plane,
{\em such that} $\hat{k_3}.\hat{z}=\mbox{cos}\,\theta $ (where, $\hat{k_3}=
\frac{\vec{k_3}}{k_3}$). 
The external gluons are lightlike, {\em viz.} $K_i^2=0$, {\em i.e.} $k_i=(k_i)_z=\epsilon$, where $\epsilon$ is the energy of the gluons and $i=1, 2, 3, 4.$
Thus the four momenta of the external gluons are written as:
\begin{align}
	K_1 (k_1^0,\vec{k_1})&\equiv K_1(\epsilon,0,0,\epsilon)\,, \qquad \qquad \qquad \qquad K_2 (k_2^0,
	\vec{k}_2)\equiv K_2(\epsilon,0,0,-\epsilon)\,\nonumber\\[0.8em]
	K_3 (k_3^0,\vec{k_3})&\equiv K_3(\epsilon,\epsilon\,\mbox{sin}\,\theta,0,\epsilon\,\mbox{cos}\,\theta)\,,\qquad \quad K_4 (k_4^0,\vec{k_4})\equiv K_4(\epsilon,-\epsilon\,\mbox{sin}\,\theta,0,-\epsilon\,\mbox{cos}\,\theta).\label{a1}
\end{align}

Since the polarisations of the external gluons are transverse 
to the respective four-momenta, so we could write 
the polarisation vectors as
\begin{align}
	\epsilon^{\mu}_{\lambda_1}(K_1)&= \frac{1}{\sqrt{2}}(0,1,\pm i,0)\,, \qquad \qquad \qquad \quad \epsilon^{\mu}_{\lambda_2}(K_2)= \frac{1}{\sqrt{2}}(0,-1,\pm i,0)
	\,\nonumber\\[0.8em]
	\epsilon^{\mu\,*}_{\lambda_3}(K_3)&= \frac{1}{\sqrt{2}}(0,-\mbox{cos}\,\theta,\pm i,\mbox{sin}\,\theta)\,,\qquad \quad \, \epsilon^{\mu\,*}_{\lambda_4}(K_4)= \frac{1}{\sqrt{2}}(0,\mbox{cos}\,\theta,\pm i,-\mbox{sin}\,\theta),\label{a2}
\end{align}
where, $\lambda_1$, $\lambda_2$ are the polarizations of incoming gluons and 
$\lambda_3$ and $\lambda_4$ are the same for the outgoing (scattered) gluons.
Since the gluons in vacuum are massless, the $\lambda_i$'s 
can only be either right-handed (R) or left-handed polarizations (L).

We begin with the $s$ channel diagram in Fig.\ref{s channel}. Using the 
Feynman rules, we write the matrix element for it as
\begin{align}
	&i\mathcal{M}_s=-ig_s^2f^{abc}f^{cde}\epsilon_{\mu}^{\lambda_1}(K_1)
	\epsilon_{\nu}^{\lambda_2}(K_2)\epsilon_{\rho}^{*\lambda_3}(K_3)
	\epsilon_{\sigma}^{*\lambda_4}(K_4)\Big[g^{\mu\nu}(K_1-K_2)^{\alpha}
	+g^{\nu\alpha}(K_1+2K_2)^{\mu}\nonumber \\&+g^{\alpha\mu}(-2K_1-K_2)^{\nu}\Big]
	\times \left(\frac{-ig_{\alpha\alpha^{\prime}}}{{(K_1+K_2)}^2}\right)
	\left[g^{\rho\sigma}(-K_3+K_4)^{\alpha^{\prime}}+
	g^{\sigma\alpha^{\prime}}(-K_3-2K_4)^{\rho}+
	g^{\alpha^{\prime}\rho}(2K_3+K_4)^{\sigma}\right],
\end{align}
which takes the form, after contracting the lorentz indices, 
\begin{align}
	i\mathcal{M}_s&=\frac{-ig_s^2f^{abc}f^{cde}}{(K_1+K_2)^2}\Big[\epsilon^{\lambda_1}(K_1)\cdot\epsilon^{\lambda_2}(K_2)\times (K_2-K_1)\cdot(K_3-K_4)\times \epsilon^{*\lambda_3}(K_3)\cdot\epsilon^{*\lambda_4}(K_4)\nonumber\\&+\epsilon^{\lambda_1}(K_1)\cdot(K_1+2K_2)\times \epsilon^{\lambda_2}(K_2)\cdot(K_4-K_3)\times \epsilon^{*\lambda_3}(K_3)\cdot\epsilon^{*\lambda_4}(K_4)-(2K_1+K_2)\cdot\epsilon^{\lambda_2}(K_2)\nonumber\\&\times (K_4-K_3)\cdot\epsilon^{\lambda_1}(K_1)\times \epsilon^{*\lambda_3}(K_3)\cdot\epsilon^{*\lambda_4}(K_4)+\epsilon^{\lambda_1}(K_1)\cdot\epsilon^{\lambda_2}(K_2)\times (K_3+2K_4)\cdot\epsilon^{*\lambda_3}(K_3)\nonumber\\&\times(K_2-K_1)\cdot\epsilon^{*\lambda_4}(K_4)-(K_1+2K_2)\cdot\epsilon^{\lambda_1}(K_1)\times (K_3+2K_4)\cdot\epsilon^{*\lambda_3}(K_3)\times \epsilon^{\lambda_2}(K_2)\cdot\epsilon^{*\lambda_4}(K_4)\nonumber\\&-\epsilon^{\lambda_1}(K_1)\cdot\epsilon^{\lambda_2}(K_2)\times \epsilon^{*\lambda_4}(K_4)\cdot(2K_3+K_4)\times (K_2-K_1)\cdot\epsilon^{*\lambda_3}(K_3)+(K_1+2K_2)\cdot\epsilon^{\lambda_1}(K_1)\nn\\&\times(2K_3+K_4)\cdot\epsilon^{*\lambda_4}(K_4)\times \epsilon^{\lambda_1}(K_1)\cdot\epsilon^{\lambda_2}(K_2)-(2K_2+K_1)\cdot\epsilon^{\lambda_2}(K_2)\times (2K_3+K_4)\cdot\epsilon^{*\lambda_4}(K_4)\nn\\&\times \epsilon^{*\lambda_3}(K_3)\cdot\epsilon^{\lambda_1}(K_1)\times+(2K_3+K_4)\cdot\epsilon^{\lambda_2}(K_2)\times (K_3+2K_4)\cdot\epsilon^{*\lambda_3}(K_3)\times \epsilon^{\lambda_1}(K_1)\cdot\epsilon^{*\lambda_4}(K_4)\Big]\label{a0}.
\end{align}
Similarly, the matrix element for the $t$-channel diagram (Fig. \ref{t channel}), 
is given by
\begin{align}
	i\mathcal{M}_t=&-ig_s^2f^{ace}f^{bde}\epsilon_{\mu}^{\lambda_1}(K_1)\epsilon_{\nu}^{\lambda_2}(K_2)\epsilon_{\rho}^{*\lambda_3}(K_3)\epsilon_{\sigma}^{*\lambda_4}(K_4)\Big[g^{\mu\rho}(K_1+K_3)^{\alpha}+g^{\rho\alpha}(K_1-2K_3)^{\mu}\nn\\&+g^{\alpha\mu}(-2K_1-K_3)^{\rho}\Big]\nonumber \times \left(\frac{-ig_{\alpha\alpha^{\prime}}}{t}\right)\left[g^{\nu\sigma}(K_4+K_2)^{\alpha^{\prime}}+g^{\sigma\alpha^{\prime}}(-K_2-2K_4)^{\nu}+g^{\alpha^{\prime}\nu}(-2K_2+K_4)^{\sigma}\right]\nn\\[0.9em]
	&=\frac{-ig_s^2f^{ace}f^{bde}}{(K_1-K_3)^2}
	\Big[(K_1+K_3) \cdot (K_2+K_4)\times 
	\epsilon^{*\lambda_1}(K_1)\cdot\epsilon^{*\lambda_3}(K_3)
	\times \epsilon^{*\lambda_3}(K_3)\cdot\epsilon^{*\lambda_4}(K_4)\nn\\&+
	\epsilon^{\lambda_2}(K_2)\cdot(K_2-2K_4)\times \epsilon^{*\lambda_4}(K_4)\cdot(K_1+K_3)\times \epsilon^{*\lambda_1}(K_1)\cdot\epsilon^{*\lambda_3}(K_3)+\epsilon^{*\lambda_2}(K_2)\cdot(K_1+K_3)\nn&\times \epsilon^{*\lambda_4}(K_4)\cdot(K_4-2K_2)\nonumber\\&\times \epsilon^{*\lambda_2}(K_2)\cdot\epsilon^{*\lambda_4}(K_4)+\epsilon^{*\lambda_1}(K_1)\cdot(K_1-2K_3)\times \epsilon^{*\lambda_3}(K_3)\cdot(K_2+K_4)\times \epsilon^{\lambda_2}(K_2)\cdot\epsilon^{*\lambda_4}(K_4)\nonumber\\&+\epsilon^{\lambda_2}(K_2)\cdot(K_1-2K_3)\times \epsilon^{\lambda_2}(K_2)\cdot(K_1-2K_4)\times \epsilon^{*\lambda_3}(K_3)\cdot\epsilon^{*\lambda_4}(K_4)+\epsilon^{*\lambda_1}(K_1)\cdot(K_1-2K_3)\nonumber\\&\times \epsilon^{*\lambda_3}(K_3)\cdot(K_4-2K_2)\times \epsilon^{\lambda_2}(K_2)\cdot\epsilon^{*\lambda_3}(K_3)+ \epsilon^{*\lambda_3}(K_3)\cdot(K_3-2K_1)\times \epsilon^{*\lambda_1}(K_1)\cdot(K_2+K_4)\nonumber\\&\times \epsilon^{*\lambda_2}(K_2)\cdot\epsilon^{*\lambda_4}(K_4)+ \epsilon^{*\lambda_1}(K_1)\cdot(K_2-2K_4)\times \epsilon^{*\lambda_3}(K_3)\cdot(K_3-2K_1)\times \epsilon^{*\lambda_1}(K_1)\cdot\epsilon^{*\lambda_4}(K_4)\nonumber\\&+\epsilon^{*\lambda_4}(K_4)\cdot(K_4-2K_2)\times \epsilon^{*\lambda_3}(K_3)\cdot(K_3-2K_1)\times \epsilon^{\lambda_1}(K_1)\cdot \epsilon^{\lambda_2}(K_2)\Big],\label{a00}
\end{align}
and the matrix element for the $u$-channel diagram in Fig \ref{u channel}
is calculated as 
\begin{align}
	i\mathcal{M}_u&=-ig_s^2f^{ade}f^{bce}\epsilon_{\mu}^{\lambda_1}(K_1)\epsilon_{\nu}^{\lambda_2}(K_2)\epsilon_{\rho}^{*\lambda_3}(K_3)\epsilon_{\sigma}^{*\lambda_4}(K_4)\Big[g^{\mu\sigma}(K_1+K_3)^{\alpha}+g^{\sigma\alpha}(K_1-2K_4)^{\mu}\nn\\&+g^{\alpha\mu}(-2K_1+K_4)^{\sigma}\Big]\times \left(\frac{-ig_{\alpha\alpha^{\prime}}}{u}\right)\left[g^{\nu\rho}(K_3+K_2)^{\alpha^{\prime}}+g^{\rho\alpha^{\prime}}(K_2-2K_4)^{\nu}+g^{\alpha^{\prime}\nu}(-2K_2+K_3)^{\rho}\right]\nn\\[0.9em]
	&=-\frac{ig_s^2f^{ade}f^{bce}}{{(K_1-K_4)}^2}\Big[(K_1+K_4)\cdot(K_3+K_2)\times \epsilon^{\lambda_1}(K_1)\cdot\epsilon^{*\lambda_4}(K_4) \times \epsilon^{\lambda_2}(K_2)\cdot\epsilon^{*\lambda_3}(K_3) \nn\\&+ \epsilon^{\lambda_2}(K_2)\cdot(K_2-2K_3) \times \epsilon^{*\lambda_3}(K_3)\cdot(K_1+K_4) \times \epsilon^{\lambda_1}(K_1)\cdot\epsilon^{*\lambda_4}(K_4) + \epsilon^{*\lambda_3}(K_3)\cdot(K_3-2K_2) \nn\\&\times \epsilon^{\lambda_2}(K_2)\cdot(K_1+K_4)  \times \epsilon^{\lambda_1}(K_1)\cdot\epsilon^{*\lambda_4}(K_4)+
	\epsilon^{\lambda_1}(K_1)\cdot(K_2-2K_4)\times\epsilon^{*\lambda_4}(K_4)\cdot(K_2+K_3)\nonumber\\&\times\epsilon^{\lambda_2}(K_2)\cdot\epsilon^{*\lambda_3}(K_3) +
	\epsilon^{\lambda_2}(K_2)\cdot(K_2-2K_3)\times\epsilon^{\lambda_1}(K_1)\cdot(K_2-2K_4)\times\epsilon^{*\lambda_3}(K_3)\epsilon^{*\lambda_4}(K_4) \nonumber\\& + 
	\epsilon^{*\lambda_3}(K_3)\cdot(K_3-2K_2)\times \epsilon^{\lambda_1}(K_1)\cdot(K_2-2K_4) \times \epsilon^{\lambda_2}(K_2)\cdot\epsilon^{*\lambda_4}(K_4)+
	\epsilon^{*\lambda_4}(K_4)\cdot(K_4-2K_1) \nonumber\\& \times \epsilon^{\lambda_1}(K_1)\cdot(K_2+K_3)\times \epsilon^{\lambda_2}(K_2)\cdot\epsilon^{*\lambda_3}(K_3) + 
	\epsilon^{\lambda_2}(K_2)\cdot(K_2-2K_3) \times \epsilon^{*\lambda_4}(K_4)\cdot(K_4-2K_1)\nonumber\\& \times \epsilon^{\lambda_1}(K_1)\cdot\epsilon^{*\lambda_3}(K_3)  + 
	\epsilon^{*\lambda_3}(K_3)\cdot(K_3-2K_2) \times\epsilon^{*\lambda_4}(K_4)\cdot(K_4-2K_1) \times \epsilon^{\lambda_1}(K_1)\cdot\epsilon^{\lambda_2}(K_2)
	\Big].\label{a000}
\end{align}
Finally, the amplitude for the 4-gluon vertex diagram (Fig. \ref{fgv}) is given by:
\begin{align}
	i\mathcal{M}_4&=-ig_s^2f^{ade}f^{bce}\epsilon_{\mu}^{\lambda_1}(K_1)\epsilon_{\nu}^{\lambda_2}(K_2)\epsilon_{\rho}^{*\lambda_3}(K_3)\epsilon_{\sigma}^{*\lambda_4}(K_4)\Big[f^{abe}f^{abe}(g^{\mu\rho}g^{\nu\sigma}-g^{\mu\sigma}g^{\nu\rho})\nonumber\\&+f^{abe}f^{abe}(g^{\mu\nu}g^{\rho\sigma}-g^{\mu\sigma}g^{\nu\rho})+f^{abe}f^{abe}(g^{\mu\nu}g^{\rho\sigma}-g^{\mu\rho}g^{\nu\sigma})\Big]\nn\\[0.9em]
	&=-ig^2\Big[f^{abc}f^{cde}\Big(\epsilon^{\lambda_1}(K_1)\cdot\epsilon^{*\lambda_3}(K_3)\times \epsilon^{\lambda_2}(K_2)\cdot\epsilon^{*\lambda_4}(K_4)-\epsilon^{*\lambda_1}(K_1)\cdot\epsilon^{*\lambda_4}(K_4)\nonumber\\&\times \epsilon^{\lambda_2}(K_2)\cdot\epsilon^{*\lambda_3}(K_3)\Big)+f^{ace}f^{bde}\Big(\epsilon^{*\lambda_1}(K_1)\cdot\epsilon^{\lambda_2}(K_2)\times \epsilon^{*\lambda_3}(K_3)\cdot\epsilon^{*\lambda_4}(K_4)-\epsilon^{*\lambda_1}(K_1)\cdot\epsilon^{*\lambda_4}(K_4)\nonumber\\&\times \epsilon^{*\lambda_2}(K_2)\cdot\epsilon^{*\lambda_3}(K_3)\Big)+f^{ade}f^{bce}\Big(\epsilon^{\lambda_1}(K_1)\cdot\epsilon^{\lambda_2}(K_2)\times \epsilon^{*\lambda_3}(K_3)\cdot\epsilon^{*\lambda_4}K_4)\nonumber\\&-\epsilon^{*\lambda_1}(K_1)\cdot\epsilon^{*\lambda_3}(K_3)\times \epsilon^{\lambda_2}(K_2)\cdot\epsilon^{*\lambda_4}(K_4) \Big) \Big].\label{a0000}
\end{align}
Since each of the polarisation indices ($\lambda_1$, $\lambda_2$, $\lambda_3$, 
$\lambda_4$) can take $R$, $L$ in \eqref{a2}, therefore the 
total possibilities of the amplitude in a scattering event will be 16, 
{\em such as}, $\mathcal{M}(RR\rightarrow RR)$, $\mathcal{M} 
(RL\rightarrow RL)$, $\mathcal{M}(LL\rightarrow LL)$, etc. 
However, the constraint of helicity conservation, reduces the 16 
possibilities to 6 only, {\em namely}
\begin{align*}
\mathcal{M}(RR\rightarrow RR)\,,\quad \mathcal{M}(LL\rightarrow LL)\,,\quad 
\mathcal{M}(RL\rightarrow RL)\,,\nonumber\\
\mathcal{M}(LR\rightarrow LR)\,,\quad \mathcal{M}(RL\rightarrow LR)\,,\quad \mathcal{M}(LR\rightarrow RL).
\end{align*} 
Moreover, the parity conservation further reduces the number of possibilities
because some processes become same, {\em viz.}
\begin{align}
i\mathcal{M}(RR\rightarrow RR)&=i\mathcal{M}(LL\rightarrow LL)\,\label{p1}\\ 
i\mathcal{M}(RL\rightarrow RL)&=i\mathcal{M}(LR\rightarrow LR)\,,\label{p2}\\
i\mathcal{M}(RL\rightarrow LR)&=i\mathcal{M}(LR\rightarrow RL).\label{p3}
\end{align}
Thus, effectively we need to calculate the matrix element only for three 
possibilities. 
Thus the amplitude for the $RR\rightarrow RR$ process is obtained by adding 
the amplitudes of all the diagrams in terms of Mandelstam variables, $s$,
$t$ and $u$
\begin{eqnarray}
i\mathcal{M}(RR\rightarrow RR) &= &i\mathcal{M}_s(RR\rightarrow RR)+
i\mathcal{M}_t(RR\rightarrow RR)+i\mathcal{M}_u(RR\rightarrow RR)\nonumber+
i\mathcal{M}_4(RR\rightarrow RR) \nonumber\\
&=&-2ig^2\left[f^{ace}f^{bde}\frac{s}{t}+f^{ade}f^{bce}\frac{s}{u}\right].\label{a51}
\end{eqnarray}
The remaining process \eqref{p1} and \eqref{p2} can be obtained by 
the following crossing symmetry: 
\begin{align}
i\mathcal{M}(RR\rightarrow RR)\xrightarrow{s\leftrightarrow u, b\leftrightarrow d}i\mathcal{M}(RL\rightarrow RL)\label{a60}\\
i\mathcal{M}(RL\rightarrow RL)\xrightarrow{u\leftrightarrow t,d\leftrightarrow c}i
\mathcal{M}(RL\rightarrow LR),\label{a6}
\end{align}
which facilitates to obatin the matrix element as
\begin{align}
i\mathcal{M}(RL\rightarrow RL)&=2ig^2\left[f^{ace}f^{bde}\frac{u}{t}+f^{abe}f^{cde}\frac{u}{s}\right]\label{a70}\\
i\mathcal{M}(RL\rightarrow LR)&=-2ig^2\left[f^{abe}f^{cde}\frac{t}{s}-
f^{ade}f^{bce}\frac{t}{u}\right].\label{a7}
\end{align}
Therefore the matrix element squared for the three processes (eq.\eqref{a51}, 
eq.\eqref{a70}, eq.\eqref{a7}), after summing over the final states is given by:
\begin{eqnarray}
\overline{|\mathcal{M}(RR\rightarrow RR)|^2}&= &
\overline{|\mathcal{M}(LL\rightarrow LL)|^2}\nonumber\\
&=& 288\,g^4\left[\frac{s^2}{t^2}+\frac{s^2}{u^2}+\frac{s^2}{ut}\right], \\
\overline{|\mathcal{M}(RL\rightarrow RL)|^2}&=&\overline{|\mathcal{M}(LR
\rightarrow LR)|^2}\nonumber\\
&=& 288\,g^4\left[\frac{u^2}{t^2}+\frac{u^2}{s^2}+\frac{u^2}{ts}\right],\\
\overline{|\mathcal{M}(RL\rightarrow LR)|^2}&=&\overline{|\mathcal{M}(LR
\rightarrow RL)|^2} \nonumber\\
&=&288\,g^4\left[\frac{t^2}{s^2}+\frac{t^2}{u^2}+\frac{t^2}{us}\right].
\end{eqnarray}

Finally, we average over the polarisations and colours in the initial state
to give rise the matrix element squared for ${\rm gg} \rightarrow {\rm gg}$ 
in the leading-order:
\begin{align}
\overline{|\mathcal{M}|^2}&=\frac{2}{8^2\times 2^2}\left[\overline{|\mathcal{M}(RR\rightarrow RR)|^2}+\overline{|\mathcal{M}(RL\rightarrow RL)|^2}+\overline{|\mathcal{M}(RL\rightarrow LR)|^2}\right]\nonumber\\[0.4em]
&=\frac{9}{2}g_s^4\left[3-\frac{ut}{s^2}-\frac{us}{t^2}-\frac{st}{u^2}\right].\label{a9}
\end{align}

\section{Different Integrals/Intermediate Steps}\label{relax}
\subsection{Differential cross section in near-forward scattering.}
Since the gluons are massless, so 
\be s+t+u=0.\label{a101}\ee 
Analyzing the scattering process in cm frame, the Mandelstam variables
become 
\begin{align}
s&=(K_1+K_2)^2\nn\\
&=4\epsilon^2\\[0.5em] t&=(K-K_3)^2\nn\\
&=-4\epsilon^2\sin^2\theta/2,
\end{align}
where, $\epsilon$ is the energy of the scattering gluons and $\theta$ is the scattering angle in the cm frame. Thus,
\begin{equation}
\left|\frac{s}{t}\right|=\frac{1}{\sin^2\theta/2}.\label{a14}
\end{equation}
In the near forward scattering, $\theta$ is close to $0$, therefore
\be 
\left(\frac{s}{t}\right)^2\gg 1\gg \frac{t}{s},\left(\frac{t}{s}\right)^2.
\label{a102}
\ee 

Therefore the matrix element squared in Eq.\eqref{a9} gets simplified
into a form
\begin{align*}
\overline{|\mathcal{M}|^2}\simeq\frac{9}{2}g_s^4\frac{s^2}{t^2}.
\end{align*}
Thus, using Eq.\eqref{D2}
\begin{align*}
\frac{d\sigma}{d\Omega}&=\frac{1}{64\pi^2s}\overline{|\mathcal{M}|^2}\nn\\&\propto\frac{1}{\sin^4\theta/2}.
\end{align*}

\subsection{Matrix element squared for resummed gluon exchange}
\begin{equation*} \mathcal{M}\propto
J_L^1\Delta_L(q_0,q)J_L^2+
\vec{J}_T^1.\vec{J}^2_T \Delta_T(q_0,q)\label{x1}\end{equation*}
where $J_L$ and $J_0$ of a general external current $J_{\mu}$ are connected 
by the current conservation via 
\be
q^{\mu}J_{\mu}=q_0J_0-qJ_L=0
\ee
and $|\vec{v}_{T}|$ and $|\vec{v_1}_{T}|$ are the transverse 
velocity components (with respect to $\vec{q}$) of the incoming gluons. Thus,
 \be 
|\vec{v}_{T}|=|v|\sin \alpha\,,\qquad |\vec{v_1}_{T}|=|v_1|
\sin \alpha. 
\ee
Hence,
\be 
\vec{v}_{T}.\vec{v_1}_{T}&=&\sin^2\alpha \cos \phi\nn\\&=&(1-x^2)
\cos \phi,
\ee
where, $\phi$ is the angle between $\vec{v}$ and $\vec{v_1}$. Thus,  the matrix element $\mathcal{M}$ becomes:
\begin{align}
\mathcal{M}&\propto g_s^2\lambda^1_{\alpha}\lambda^2_{\alpha}E\,E_1\Delta_L(q_0,q)+g_s^2\lambda^1_{\alpha}\lambda^2_{\alpha}E\,E_1(1-x^2)\cos \phi\Delta_T(q_0,q)\nn\\[0.4em]
&\propto \Delta_L(q_0,q)+(1-x^2)\cos\phi \,\Delta_T(q_0,q) \nn\\
&=A (k,k^\prime) \left( \Delta_L(q_0,q)+(1-x^2)\cos\phi \,\Delta_T(q_0,q)
\right),
\label{x2}
\end{align}
which, in the limit of $m_{gT} \rightarrow 0$, reduces to 
\begin{align}
\overline{|\mathcal{M}|^2}&=A^2\left[-\frac{1}{q^2}+(1-x^2)\,\mbox{cos}\,\phi\, \frac{1}{q^2-q_0^2}\right]^2\nn\\
&=A^2\frac{(1-\mbox{cos}\,\phi)^2}{q^4}.
\label{a103}
\end{align}
The function, $A(k,k^\prime)$ could be fixed by the condition,
where the matrix element squared in \eqref{a103} 
should reduce to the matrix element squared calculated in vacuum:
\begin{equation*}
\overline{|\mathcal{M}|^2} = 18g^4\frac{k^2\,k^{\prime\,2}}{q^4}
(1-\mbox{cos}\,\phi)^2.
\end{equation*}
Thus, $A(k,k^\prime)$ has been obtained as
\be
A^2=18g^4k^2k^{\prime\,2},
\ee
hence the matrix element squared with resummed gluon propagator 
becomes
\begin{eqnarray*}
\overline{|\mathcal{M}|^2}=18g^4k^2k^{\prime\, 2}|\Delta_L(q_0,q)+(1-x^2)\mbox{cos}\,\phi\, \Delta_T(q_0,q)|^2.\label{h}
\end{eqnarray*}

\subsection{Derivation of Eq.(26)}
The collision term is given by:
\begin{eqnarray*}
C[f] &=& \frac{\nu_g}{2 E_{\vec{p}}}\int 
\frac{d^3 p_1}{{(2\pi)}^3 2 E_{\vec{p}_1}}
\frac{d^3 p^\prime}{{(2\pi)}^3 2 E_{\vec{p}^\prime}}
\frac{d^3 p_1^\prime}{{(2\pi)}^3 2 E_{\vec{p}_1^\prime}} 
(2\pi)^4\delta^{(4)}(P+P_1-P^{\prime}-P_1^{\prime}) \nonumber\\
&&\left[f^{\prime}f_1^{\prime}(1+f)(1+f_1)-ff_1(1+f^{\prime})(1+f_1^{\prime})\right]\overline{|\mathcal{M}|^2}.\label{x5}
\end{eqnarray*}
\begin{multline}
f^{\prime}f_1^{\prime}(1+f)(1+f_1)=f^{\prime\,(0)}\left[1-\beta(1+f^{\prime\,(0)})\,\Gamma^{\prime}\,\frac{\partial u_x}{\partial y}\right]f_1^{\prime\,(0)}\left[1-\beta(1+f_1^{\prime\,(0)})\,\Gamma_1^{\prime}\,\frac{\partial u_x}{\partial y}\right]\\
\left\{1+f^{(0)}\left[1-\beta(1+f^{(0)})\,\Gamma\,\frac{\partial u_x}{\partial y}\right]\right\}\times \left\{1+f_1^{(0)}\left[1-\beta(1+f_1^{(0)})\,\Gamma_1^{\prime}\,\frac{\partial u_x}{\partial y}\right]\right\}.\label{o}
\end{multline}
The relevant terms in the above product are the ones linear in $\frac{\partial u_x}{\partial y}$:
\begin{multline*}
-\Bigg[f^{\prime\,(0)}f_1^{\prime\,(0)}(1+f_1^{\prime\,(0)})\beta\, \Gamma_1^{\prime}\frac{\partial u_x}{\partial y}+f^{\prime\,(0)}f_1^{\prime\,(0)}(1+f^{\prime\,(0)})\beta\, \Gamma^{\prime}\frac{\partial u_x}{\partial y}+f^{\prime\,(0)}f_1^{\prime\,(0)}(1+f_1^{\prime\,(0)})\beta\, \Gamma_1^{\prime}f_1^{(0)}\frac{\partial u_x}{\partial y}+\\
f^{\prime\,(0)}f_1^{\prime\,(0)}(1+f^{\prime\,(0)})\beta\, \Gamma^{\prime}f_1^{(0)}\frac{\partial u_x}{\partial y}+
f^{\prime\,(0)}f_1^{\prime\,(0)}(1+f_1^{\prime\,(0)})\beta\, \Gamma_1^{\prime}f^{(0)}\frac{\partial u_x}{\partial y}+f^{\prime\,(0)}f_1^{\prime\,(0)}(1+f_1^{\prime\,(0)})\beta\, \Gamma_1^{\prime}f^{(0)}f_1^{(0)}\frac{\partial u_x}{\partial y}+\\
f^{\prime\,(0)}f_1^{\prime\,(0)}(1+f^{\prime\,(0)})\beta\, \Gamma^{\prime}f^{(0)}f_1^{(0)}\frac{\partial u_x}{\partial y}+f^{\prime\,(0)}f_1^{\prime\,(0)}f^{(0)}(1+f^{(0)})\beta\, \Gamma\frac{\partial u_x}{\partial y}+f^{\prime\,(0)}f_1^{\prime\,(0)}f_1^{(0)}(1+f_1^{(0)})\beta\, \Gamma_1\frac{\partial u_x}{\partial y}\\
+f^{\prime\,(0)}f_1^{\prime\,(0)}f^{(0)}f_1^{(0)}(1+f_1^{(0)})\beta\, \Gamma_1\frac{\partial u_x}{\partial y}+f^{\prime\,(0)}f_1^{\prime\,(0)}f^{(0)}f_1^{(0)}(1+f^{(0)})\beta\, \Gamma\frac{\partial u_x}{\partial y}\Bigg].
\end{multline*}
\begin{multline*}
=-\beta(1+f^{\prime\,(0)})(1+f_1^{\prime\,(0)})f^{(0)}f_1^{(0)}\frac{\partial u_x}{\partial y}\Bigg[\Gamma_1^{\prime}\frac{f^{\prime\,(0)}f_1^{\prime\,(0)}}{1+f^{\prime\,(0)}}\frac{(1+f^{(0)})(1+f_1^{(0)})}{f^{(0)}f_1^{(0)}}+\\
\Gamma^{\prime}\frac{f^{\prime\,(0)}f_1^{\prime\,(0)}}{1+f_1^{\prime\,(0)}}\frac{(1+f^{(0)})(1+f_1^{(0)})}{f^{(0)}f_1^{(0)}}+\Gamma\frac{f^{\prime\,(0)}f_1^{\prime\,(0)}(1+f^{(0)})(1+f_1^{(0)})}{(1+f^{\prime\,(0)})(1+f_1^{\prime\,(0)})f_1^{(0)}}+\Gamma_1\frac{f^{\prime\,(0)}f_1^{\prime\,(0)}(1+f^{(0)})(1+f_1^{(0)})}{(1+f^{\prime\,(0)})(1+f_1^{\prime\,(0)})f^{(0)}}\Bigg].
\end{multline*}
 We finally arrive at
\begin{equation*}
f^{\prime}f_1^{\prime}(1+f)(1+f_1)=-\beta(1+f^{\prime\,(0)})(1+f_1^{\prime\,(0)})f^{(0)}f_1^{(0)}\frac{\partial u_x}{\partial y}\left[\Gamma_1^{\prime}(1+f_1^{\prime\,(0)})+\Gamma^{\prime}(1+f^{\prime\,(0)})+\Gamma f^{(0)}+\Gamma_1 f_1^{(0)}\right].
\end{equation*}
Similarly, it can be shown that
\begin{equation*}
ff_1(1+f^{\prime})(1+f_1^{\prime})=-\beta(1+f^{\prime\,(0)})(1+f_1^{\prime\,(0)})f^{(0)}f_1^{(0)}\frac{\partial u_x}{\partial y}\left[\Gamma_1^{\prime}f_1^{\prime\,(0)}+\Gamma^{\prime}f^{\prime\,(0)}+\Gamma (1+f^{(0)})+\Gamma_1 (1+f_1^{(0)})\right].
\end{equation*}
Thus,
\begin{align}
 f^{\prime}f_1^{\prime}(1+f)(1+f_1)-ff_1(1+f^{\prime})(1+f_1^{\prime})&=\beta (1+f^{\prime\,(0)})(1+f_1^{\prime\,(0)})f^{(0)}f_1^{(0)}\frac{\partial u_x}{\partial y}\left[\Gamma+\Gamma_1-\Gamma^{\prime}-\Gamma_1^{\prime}\right]\nn\\
&\simeq \beta (1+n^{\prime})(1+n_1^{\prime})\,n\,n_1\frac{\partial u_x}{\partial y}\left[\Gamma+\Gamma_1-\Gamma^{\prime}-\Gamma_1^{\prime}\right]. \label{x3}
\end{align}
\subsection{Derivation of Eq.(30)}
\begin{equation*}
\eta = - \nu_g\int \frac{\mbox{d}^3p}{(2\pi)^3}  
p_{x}v_{y}\frac{\partial n}{\partial E}\,\Gamma.\label{r}
\end{equation*}
\begin{align*}
\beta p_xv_y=&\frac{\nu_g}{2E}\int \frac{\mbox{d}^3p}{(2\pi_1)^3} \, \frac{\mbox{d}^3p}{(2\pi^{\prime})^3} \,\frac{\mbox{d}^3p_1^{\prime}}{(2\pi)^3} (2\pi)^4\delta^{(4)}(P+P_1-P^{\prime}-P_1^{\prime})\times \overline{|\mathcal{M}|^2}\nn\\
&\frac{n_1(1+n^{\prime})(1+n_1^{\prime})}{1+n}\left[\Gamma+\Gamma_1-\Gamma^{\prime}-\Gamma_1^{\prime}\right].
\end{align*}
Thus,
\begin{align}
\eta&=\nu_g^2\beta\int \frac{\mbox{d}^3p}{(2\pi)^3}\,\frac{\mbox{d}^3p_1}{(2\pi)^3}\,\frac{\mbox{d}^3p^{\prime}}{(2\pi)^3}\,\frac{\mbox{d}^3p_1^{\prime}}{(2\pi)^3}\,n\,n_1(1+n^{\prime})(1+n_1^{\prime}) (2\pi)^4\times\nn\\
&\delta^{(4)}(P+P_1-P^{\prime}-P_1^{\prime})\times \overline{|\mathcal{M}|^2}\left[\Gamma+\Gamma_1-\Gamma^{\prime}-\Gamma_1^{\prime}\right]\Gamma.\label{s}
\end{align}
The integration is over all the variables $p$, $p_1$, $p^{\prime}$, $p_1^{\prime}$ and so, it is possible to rename the variables without altering the value of the integral.

Effecting $p\leftrightarrow p_1$,\quad $p^{\prime}\leftrightarrow p_1^{\prime}$, we get
\begin{align}
\eta=&\nu_g^2\beta\int \frac{\mbox{d}^3p}{(2\pi)^3}\,\frac{\mbox{d}^3p_1}{(2\pi)^3}\,\frac{\mbox{d}^3p^{\prime}}{(2\pi)^3}\,\frac{\mbox{d}^3p_1^{\prime}}{(2\pi)^3}\,n\,n_1(1+n^{\prime})(1+n_1^{\prime}) (2\pi)^4\times\nn\\
&\delta^{(4)}(P+P_1-P^{\prime}-P_1^{\prime})\times \overline{|\mathcal{M}|^2}\left[\Gamma+\Gamma_1-\Gamma^{\prime}-\Gamma_1^{\prime}\right]\Gamma_1.\label{t}
\end{align}
From eq.\eqref{t} and eq.\eqref{s}, we have,
\begin{align}
\eta=&\frac{\nu_g^2\beta}{2}\int\frac{\mbox{d}^3p}{(2\pi)^3}\,\frac{\mbox{d}^3p_1}{(2\pi)^3}\,\frac{\mbox{d}^3p^{\prime}}{(2\pi)^3}\,\frac{\mbox{d}^3p_1^{\prime}}{(2\pi)^3}\,n\,n_1(1+n^{\prime})(1+n_1^{\prime}) (2\pi)^4\times\nn\\
&\delta^{(4)}(P+P_1-P^{\prime}-P_1^{\prime})\times \overline{|\mathcal{M}|^2}\left[\Gamma+\Gamma_1-\Gamma^{\prime}-\Gamma_1^{\prime}\right][\Gamma+\Gamma_1].\label{u}
\end{align}
Again, we effect $p_1\leftrightarrow p^{\prime}$,\quad $p\leftrightarrow p_1^{\prime}$, to get
\begin{align}
\eta=&\frac{\nu_g^2\beta}{2}\int\frac{\mbox{d}^3p}{(2\pi)^3}\,\frac{\mbox{d}^3p_1}{(2\pi)^3}\,\frac{\mbox{d}^3p^{\prime}}{(2\pi)^3}\,\frac{\mbox{d}^3p_1^{\prime}}{(2\pi)^3}\,n^{\prime}n_1^{\prime}(1+n)(1+n_1) (2\pi)^4\times\nn\\
&\delta^{(4)}(P^{\prime}+P_1^{\prime}-P-P_1)\times \overline{|\mathcal{M}|^2}\left[-(\Gamma+\Gamma_1-\Gamma^{\prime}-\Gamma_1^{\prime})\right][\Gamma^{\prime}+\Gamma_1^{\prime}]\nn\\[0.6em]
=&\frac{\nu_g^2\beta}{2}\int \frac{\mbox{d}^3p}{(2\pi)^3}\,\frac{\mbox{d}^3p_1}{(2\pi)^3}\,\frac{\mbox{d}^3p^{\prime}}{(2\pi)^3}\,\frac{\mbox{d}^3p_1^{\prime}}{(2\pi)^3}\,n\,n_1(1+n^{\prime})(1+n_1^{\prime}) (2\pi)^4\times\nn\\
&\delta^{(4)}(P+P_1-P^{\prime}-P_1^{\prime})\times \overline{|\mathcal{M}|^2}\left[-(\Gamma+\Gamma_1-\Gamma^{\prime}-\Gamma_1^{\prime})\right][\Gamma^{\prime}+\Gamma_1^{\prime}].\label{w}
\end{align}
Finally,
\begin{align}
\eta=&\frac{\nu_g^2\beta}{2}\int\frac{\mbox{d}^3p}{(2\pi)^3}\,\frac{\mbox{d}^3p_1}{(2\pi)^3}\,\frac{\mbox{d}^3p^{\prime}}{(2\pi)^3}\,\frac{\mbox{d}^3p_1^{\prime}}{(2\pi)^3}\,n\,n_1(1+n^{\prime})(1+n_1^{\prime}) (2\pi)^4\times\nn\\
&\delta^{(4)}(P+P_1-P^{\prime}-P_1^{\prime})\times \overline{|\mathcal{M}|^2}\left[\Gamma+\Gamma_1-\Gamma^{\prime}-\Gamma_1^{\prime}\right][\Gamma+\Gamma_1-\Gamma^{\prime}-\Gamma_1^{\prime}]^2.\label{x}
\end{align}
Using eq.\eqref{R} and eq.\eqref{x}, a convenient formula for $\eta$ is:
\begin{align*}
\frac{1}{\eta}=&
\Bigg(\frac{\beta}{4}\int \frac{\mbox{d}^3p}{(2\pi)^3}\,\frac{\mbox{d}^3p_1}{(2\pi)^3}\,\frac{\mbox{d}^3p^{\prime}}{(2\pi)^3}\,\frac{\mbox{d}^3p_1^{\prime}}{(2\pi)^3}\, n\,n_1(1+n^{\prime})(1+n_1^{\prime})\,\overline{|\mathcal{M}|^2}\times (2\pi)^4\nn\\
&\left. \delta^{(4)}(P+P_1-P^{\prime}-P_1^{\prime})[\Gamma+\Gamma_1-\Gamma^{\prime}-
\Gamma_1^{\prime}]^2\Bigg) \middle/  \left(\int \frac{d^3p}{(2\pi)^3}p_{x}v_{y}
\frac{\partial n}{\partial E}\,\Gamma\right)^{2}\right..
\end{align*}
\subsection{Integral in denominator of Eq.(30)}
Instead of $\Gamma$, let us choose a function $\Psi=p\,p_xv_y$, and solve for $\eta$. Then, we have:
\begin{align*}
p_x v_y\frac{\mbox{d}n}{\mbox{d}E}\,\Psi&=p_xv_y\frac{\mbox{d}n}{\mbox{d}E}\,p\,p_xv_y\\
&=p_x^2\,\frac{p_y}{E}\frac{\mbox{d}n}{\mbox{d}p}\,p_y\\
&=p_x^2\,p_y^2\,\frac{\mbox{d}n}{\mbox{d}p}\frac{1}{p},
\end{align*}
where, we have used the fact that the 4-momentum of an external gluon is lightlike. With this, the integral in the denominator of Eq.\eqref{Y} becomes:
\begin{equation}
\int \frac{\mbox{d}^3p}{(2\pi)^3}p_x^2\,p_y^2\,\frac{1}{p}\frac{\mbox{d}n}{\mbox{d}p}.
\end{equation}
This is a standard integral and evaluates to:
\begin{equation*}
\int \frac{\mbox{d}^3p}{(2\pi)^3}p_x^2\,p_y^2\,\frac{1}{p}\frac{\mbox{d}n}{\mbox{d}p}=-\frac{4T^5\zeta(5)}{\pi^2}.
\end{equation*}
\subsection{Integral in numerator of Eq.(30)}
Next, to evaluate the integral in the the numerator of Eq.\eqref{Y}, we change the integration variables to $\vec{k}$, $\vec{k^{\prime}}$ and $\vec{q}$. From Eq.\eqref{H}, it is evident that $\overline{|\mathcal{M}|^2}$ depends only on the magnitudes and relative orientations of $\vec{k}$, $\vec{k^{\prime}}$ and $\vec{q}$.
By fixing the magnitudes and relative orientations of $\vec{k}$, $\vec{k^{\prime}}$ and $\vec{q}$,  $[\Psi+\Psi_1-\Psi^{\prime}-\Psi_1^{\prime}]^2$ is averaged  over the three euler angles between the aforementioned vectors and a fixed reference frame. For two arbitrary vectors $\vec{X}$ and $\vec{Y}$, the angular average is given by:
\begin{equation}
\langle X_iX_jY_kY_l\rangle=C_1\delta_{ij}\delta_{kl}+C_2(\delta_{ik}\delta_{jl}+\delta_{il}\delta_{jk}).
\end{equation}
with
$$C_1=\frac{1}{15}\left[2\vec{X}^2\,\vec{Y}^2-(\vec{X}.\vec{Y})^2\right]\,,\qquad C_2=\frac{1}{30}\left[-\vec{X}^2\,\vec{Y}^2+3(\vec{X}.\vec{Y})^2\right]. $$
Thus, the angular averages evaluates to
\begin{align}
\langle [\Psi+\Psi_1-\Psi^{\prime}-\Psi_1^{\prime}]^2\rangle &=q^2W(k,k^{\prime};x,\phi)\nonumber \\
&=\frac{q^2}{15}\left[3(k^2+k^{\prime\,2})+x^2(k-k^{\prime})^2-kk^{\prime}(\hat{k}.\hat{k^{\prime}})\right].\label{z1}
\end{align}
We now simplify the Boltzmann factors in the numerator term of eq.\eqref{Y}. From eq.\eqref{c}, we have
\begin{equation}
 E =k+q^0/2\,, \qquad E^{\prime}=k-q^0/2.
\end{equation}
Thus,
$$n(1+n^{\prime})=n(k+q^0/2)\left\{1+n(k-q^0/2)\right\}.$$
We make use of the relation $f(k^0)f(q^0-k^0)=f(q^0)+f(q^0)f(k^0)+f(q^0)f(q^0-k^0)$ to write:
\begin{equation}
n(1+n^{\prime})=f(q^0)\left\{f(k-q^0/2)-f(k+q^0/2)\right\}.
\end{equation}
Taylor expanding, we get:
\begin{equation}
n(1+n^{\prime})=-q^0\frac{\mbox{d}n}{\mbox{d}k}f(q^0).\label{z2}
\end{equation}
Similarly, it can be shown that
\begin{equation}
n_1(1+n_1^{\prime})=q^0\frac{\mbox{d}n}{\mbox{d}k^{\prime}}f(-q^0).\label{z3}
\end{equation}
With the new integration variables, the phase space factor becomes:
\begin{equation}
\int \mbox{d}^3p\cdots\mbox{d}^3p_1^{\prime}(2\pi)^4
\delta^{(4)}(P+P_1-P^{\prime}-P_1^{\prime})=2\times (2\pi)^7\int q\,\mbox{d}q\,k^2\mbox{d}k\,k^{\prime\,2}\mbox{d}k^{\prime}\mbox{d}x\frac{\mbox{d}\phi}{2\pi}.\label{phase}
\end{equation}
Substituting eq.\eqref{phase}, eq.\eqref{z3}, eq.\eqref{z2}, eq.\eqref{z1}, in eq.\eqref{Y}, the numerator term becomes:
\begin{align}
N=&\frac{18 \beta g^4}{32(2\pi)^5}\int k^2\frac{\mbox{d}n}{\mbox{d}k}\mbox{d}k\int k^{\prime\,2}\frac{\mbox{d}n}{\mbox{d}k^{\prime}}\mbox{d}k^{\prime}\int \mbox{d}x\int \frac{\mbox{d}\phi}{2\pi}\int \mbox{d}q\, q^3(qx)^2f(qx)\nn\\
&[1+f(qx)]W(k,k^{\prime};x,\phi)\times |\Delta_L(q_0,q)+(1-x^2)\mbox{cos}\,\phi\, \Delta_T(q_0,q)|^2. \label{z4}
\end{align}
Using the approximation $(qx)^2f(qx)[1+f(qx)]\simeq T^2$ for $q\ll T$, the $x\mbox{-}q$ integral is written as
\begin{equation}
\int_{-1}^{1} \mbox{d}x\int \mbox{d}q\, q^3|\Delta_L(q_0,q)+(1-x^2)\mbox{cos}\,\phi\, \Delta_T(q_0,q)|^2.
\end{equation}
The only possible source of divergence in the above integration is from $| \Delta_T(q_0,q)|^2$ in the region $x\rightarrow 0$. In that limit, retaining the leading term in $x$ leads to the following expression of $\Delta_T$:
\begin{equation}
\Delta_T(q_0,q)\simeq \frac{1}{q^2-i\pi m_{gT}^2x/2}.
\end{equation}
So we investigate the contribution of $| \Delta_T(q_0,q)|^2$ in the above integral. Since the $q$ integration is cut-off at $q_{max}\sim T$ by the Bose-einstein factors, we write:
\begin{equation}
\int_{0}^{T} \left| \Delta_T(q_0,q)\right|^2=\int_{0}^{T}\frac{q^3\,\mbox{d}q}{|q^2-i\pi m_{gT}^2x/2|^2}.
\end{equation}

	
	$$\left|q^{2}-i\pi m_{gT}^{2}\frac{x}{2}\right|^2 = \left(q^{2}-i\pi m_{gT}^{2}\frac{x}{2}\right)\left(q^{2}+i\pi m_{gT}^{2}\frac{x}{2}\right)$$
	$$= \left(q^{4}+\pi^{2} m_{gT}^{4}\frac{x^2}{4}\right)$$
	So, the integral becomes
	$$I=\int_{0}^{T} \frac{q^{3}}{q^{4}+\pi^{2} m_{gT}^{4}\frac{x^2}{4}}dq.$$

	With the substitution  $q^{4}+\pi^{2} m_{gT}^{4}\frac{x^2}{4}=u$, the integral becomes:
	\begin{equation*}
	\begin{split}
	I&=\frac{1}{4}\int_{a}^{b}\frac{du}{u}\,;\qquad\mbox{with}\,\, a=\pi^{2} m_{gT}^{4}\frac{x^2}{4}\,,\qquad b=T^4+\pi^{2} m_{gT}^{4}\frac{x^2}{4}\\
	&=\frac{1}{4}\left[\ln(b)-\ln(a)\right]\\
	&=\frac{1}{4}\ln\left(\pi^{2}\frac{x^2}{4}+\frac{T^4}{m_{gT}^4}\right)-\frac{1}{4}\ln\left(\frac{\pi x}{2}\right)^2. 
	\end{split}
	\end{equation*}
	We can put $x = 0$ in the first logarithm, to get,
	$$ I= \frac{1}{4}\ln\left(\frac{T}{m_{gT}}\right)^4-\frac{1}{2}\ln\left(\frac{\pi x}{2}\right).$$
	Thus,
	$$\int_{0}^{T}\frac{q^{3}}{|q^{2}-i\pi m_{gT}^{2}\frac{x}{2}|^2}dq = \ln\left(\frac{T}{m_{gT}}\right)-\frac{1}{2}\ln\left(\frac{\pi x}{2}\right).$$
Thus, transverse gluon exchange is screened by the thermal gluon mass $m_{gT}$. The contribution of the interference term can be evaluated similarly and in the leading logarithmic approximation, we have:
\begin{equation}
\int \mbox{d}q\,q^3|\Delta_L(q_0,q)+(1-x^2)\mbox{cos}\,\phi\, \Delta_T(q_0,q)|^2=(1-\mbox{cos}\,\phi)^2\,\mbox{ln}\left(\frac{T}{m_{gT}}\right).\label{z5}
\end{equation}
Next, we evaluate the $x$ and $\phi$ integrals.
$$\int_{-1}^{1} \mbox{d}x\int_{0}^{2\pi}\frac{\mbox{d}\phi}{2\pi}\,W(k,k^{\prime};x,\phi)(1-\mbox{cos}\,\phi)^2$$
\begin{align*}
=&\int_{-1}^{1} \mbox{d}x\int_{0}^{2\pi}\frac{\mbox{d}\phi}{2\pi}\,\frac{k^2+k^{\prime\,2}}{5}(1-\mbox{cos}\,\phi)^2+\int_{-1}^{1} \mbox{d}x\int_{0}^{2\pi}\frac{\mbox{d}\phi}{2\pi}\,\frac{x^2(k-k^{\prime})^2}{15}(1-\mbox{cos}\,\phi)^2-\\
&\int_{-1}^{1} \mbox{d}x\int_{0}^{2\pi}\frac{\mbox{d}\phi}{2\pi}\,\frac{1}{15}k\,k^{\prime}\mbox{cos}\,\phi(1-\mbox{cos}\,\phi)^2.
\end{align*}

The three integrals above, evaluate to $\frac{3}{5}(k^2+k^{\prime\,2})$, $\frac{(k-k^{\prime})^2}{15}$ and $\frac{-2}{15}\,kk^{\prime}$ respectively. Thus,
\begin{align}
\int_{-1}^{1} \mbox{d}x\int_{0}^{2\pi}\frac{\mbox{d}\phi}{2\pi}\,W(k,k^{\prime};x,\phi)(1-\mbox{cos}\,\phi)^2=\frac{2}{3}(k^2+k^{\prime\,2}).\label{z6}
\end{align}
Finally, we are left with the $k\mbox{-}k^{\prime}$ integral:
\begin{equation}
\frac{2}{3}\int_{0}^{\infty}k^2\,\frac{\mbox{d}n}{\mbox{d}k}\, k^{\prime\,2}\frac{\mbox{d}n}{\mbox{d}k^{\prime}}\,(k^2+k^{\prime\,2})\,\mbox{d}k\,\mbox{d}k^{\prime}=\frac{2}{3}\times 2\times \int_{0}^{\infty}k^4\,\frac{\mbox{d}n}{\mbox{d}k}\,\mbox{d}k\,k^{\prime\,2}\frac{\mbox{d}n}{\mbox{d}k^{\prime}}\,\mbox{d}k^{\prime}.
\end{equation}
Now,
\begin{align*}
\int_{0}^{\infty}k^4\,\frac{\mbox{d}n}{\mbox{d}k}\,\mbox{d}k&=-4\,T^4\,\zeta(4)\Gamma(4)\,;\qquad \Gamma \equiv \mbox{Gamma function}\\
&=-\frac{4}{15}T^4\,\pi^4.
\end{align*}
Similarly,
\begin{align*}
\int_{0}^{\infty}k^{\prime\,2}\frac{\mbox{d}n}{\mbox{d}k^{\prime}}\,\mbox{d}k^{\prime}=-\frac{\pi^2\,T^2}{3}.
\end{align*}
Thus, the $k\mbox{-}k^{\prime}$ integral evaluates to: 
\begin{equation*}
\frac{4}{3}\times \frac{4}{15}T^4\,\pi^4\times \frac{\pi^2\,T^2}{3}=\frac{16}{3\times 45}\pi^6\,T^6.
\end{equation*}
Collating all the results, the numerator term comes out to be:
\begin{equation*}
N=\frac{\pi^3T^7\alpha_s^2}{30}.
\end{equation*}
\subsection{Boltzmann transport equation in relaxation time approximation}
The relaxation time approximation of the Boltzmann transport equation is defined as:
\begin{align} Df&=-\frac{f-f^{(0)}}{\tau}\nn\\
&=-\frac{\delta f}{\tau}\label{z9},
\end{align}
where, $\tau$ is the relaxation time. As already mentioned in the text,
\begin{equation*} Df= -p_xv_y\frac{\partial f^{(0)}}{\partial E}\frac{\partial u_x}{\partial y}\label{z10},\end{equation*}
and,
\begin{equation*} \delta f=\frac{\partial f^{(0)}}{\partial E}\Gamma(p)\frac{\partial u_x}{\partial y}.\end{equation*}
Thus, eq.\eqref{z9} reduces to:
\begin{equation*} p_xv_y=\frac{\Gamma(p)}{\tau}.
\end{equation*}
\subsection{Derivation of eq.(38)}
Using Eq.\eqref{z11}, Eq.\eqref{R} becomes
\begin{align}
\eta = - \nu_g\tau\int \frac{\mbox{d}^3p}{(2\pi)^3}  
p_{x}^2v_{y}^2\frac{\partial n}{\partial E}.
\end{align}
\begin{align}
i.e.\quad \frac{1}{\tau}&=\frac{-\nu_g}{\eta}\int \frac{\mbox{d}^3p}{(2\pi)^3}  
p_{x}^2v_{y}^2\frac{\partial n}{\partial E}\nn\\
&=\frac{-\nu_g}{\eta (2\pi)^3}\int \mbox{d}p\,p^4\frac{\mbox{d}n}{\mbox{d}p}\int \sin^4\theta^{\prime}\cos^2\phi^{\prime}\sin^2\phi^{\prime}\,\mbox{d}\Omega^{\prime},
\end{align}
where, $\theta^{\prime}$ and $\phi^{\prime}$ are the polar and azimuthal angles respectively, with $\mbox{d}\Omega^{\prime}=\sin\theta^{\prime}\,\mbox{d}\theta^{\prime}\,\mbox{d}\phi^{\prime}$. The radial integral has already been evaluated earlier and its value is $\frac{-4}{15}\pi^4T^4$. The angular integral evaluates to $\frac{4\pi}{15}$. Thus,
\begin{align*}
\frac{1}{\tau}&=\frac{32}{225}\pi^2\frac{T^4}{\eta}\nn\\
&\simeq 1.404\frac{T^4}{\eta}.
\end{align*}

\end{appendices}

\end{document}